\documentclass[aps,pre,twocolumn]{revtex4}
\pdfoutput=1
\usepackage{graphicx}
\usepackage{amssymb,amsfonts,amsmath}

\begin{document}

\title{Harmonic Order Parameters for Characterizing Complex Particle Morphologies}


\author{Aaron S. Keys$^1$}
\author{Christopher R. Iacovella$^1$}
\author{Sharon C. Glotzer$^{1,2}$}
\affiliation{$^1$Department of Chemical Engineering and $^2$Department of Materials Science and Engineering \\University of Michigan, Ann Arbor, Michigan 48109-2136}

\date{\today}

\begin{abstract}

Order parameters based on spherical harmonics and Fourier coefficients already play a significant role in condensed matter research in the context of systems of spherical or point particles.  Here, we extend these types of order parameter to more complex shapes, such as those encountered in nanoscale self-assembly applications.  To do so, we build on a powerful set of techniques that originate in the computer science field of ``shape matching.''  We demonstrate how shape matching techniques can be applied to identify unknown structures and create highly-specialized \textit{ad hoc} order parameters.  Additionally, we investigate the special symmetry properties of harmonic descriptors, and demonstrate how they can be exploited to provide optimal solutions to certain classes of problems.  Our techniques can be applied to particle systems in general, both simulated and experimental, provided the particle positions are known.  

\end{abstract}


\maketitle


Quantitative measures of symmetry and order, such as order parameters and correlation functions, are often applied within the physical and chemical sciences to study the structural properties of particle systems.  Structural quantities are particularly important in condensed systems, including nano and colloidal scale self-assembly applications, where subtle differences in particle ordering can greatly affect the thermodynamical, physical, chemical, electrical, and optical properties of a material or device\cite{meyers2008, wang2005, yoon2005, confinedfluids, murray, kumacheva, glotzer07}.  Some particularly useful order parameters, known as  ``bond order parameters,'' were introduced by Nelson and co-workers in the context of 2d and 3d simulations of point particles\cite{halperin78, snr83}.  These order parameters have since been widely applied to both simulated and experimental systems for quantifying crystal-like ordering in systems of spherical particles.  Some common applications of bond order parameters include, but are not limited to, identifying small ordered clusters\cite{snr83, gasser, iac07},  constructing static\cite{marcus96, kawasaki, ernstnagelgrest, tenwolde96, auer04, gasser01} and temporal\cite{tanaka, keys07} correlation functions, identifying structural defects\cite{tesfuv2}, and studying nucleation and growth\cite{tenwolde96, auer04}. Bond order parameters have the advantage that they can give a representative value for a given structure regardless of spatial orientation.  Additionally, they are robust under random perturbations due to thermal noise and highlight important rotational symmetries.  

Since bond order parameters were originally designed to quantify order in small point clusters\cite{halperin78, snr83}, they cannot be applied directly to complex structures or particle shapes.  Thus, they fail, in many cases, to fully describe the complex structures that arise in contemporary disciplines, such as nanoscale and colloidal assembly, soft matter physics, and the biological sciences.  The field of nanoparticle assembly in particular encompasses a vast range of the structural complexity that is possible for particle systems\cite{glotzer07, glotzer2005, kumacheva,desimone2006}.  Here, nanometer and micron-sized colloidal particles with a wide range of shapes, compositions and interparticle forces self-assemble into unique structures, such as complex crystals reminiscent of atomistic condensed matter\cite{amir09, tang2006, zhang2007, zhang2003}, phase-separating structures similar to those observed for block copolymer and surfactant systems\cite{glotzer2005, zhang2003, park, reister, arthi2008, waddon2002, rotello}, and hierarchical assemblies that resemble certain biological structures\cite{virus, hagan2006}.  Because there are no standard accepted structural metrics or order parameters for describing these systems, \textit{ad hoc} analyses or visual inspection are often employed instead.  This often yields relatively inaccurate or incomplete results when compared to statistical analysis.

Here, we generalize and extend bond order parameters for applications involving complex structures.  Our primary focus is on the field of self-assembly, where we predict that these new structural characterization schemes are immediately required and readily applicable.  However, the structural metrics that we introduce, known as ``harmonic descriptors,'' can be applied across diverse fields, including soft matter, macromolecular and biological sciences, and other applications where complex particle structures are encountered.  To derive general structural metrics for these systems, we draw strongly from the computer science field of shape matching\cite{keys10-ARCMP, keys10-LONG}, which aids us both in constructing mathematical representations for describing complex structures, and in quantifying structural similarity based on these representations.  Our approach involves constructing ``shape descriptors'' based on a collection of many bond order parameters computed over a range of length scales.  By quantifying the similarity between pairs of these shape descriptors, we can derive order parameters and correlation functions that are applicable to complex structures.  In this respect, the order parameters that we introduce actually represent the degree to which two collections of ``sub-order parameters'' match.  This pairwise comparative approach to constructing order parameters is necessary for applications involving complex structures, which, unlike simple clusters of spherical particles, require a range of structural metrics to describe them completely.  Shape matching techniques based on the harmonic descriptors described here have already been applied to complex particle systems in the context of fast database searches for retrieving macromolecules and proteins\cite{yeh, mak, venkatraman}.  We have applied similar database searches in the context of characterizing local structure in nanoparticle assemblies\cite{iac07, amir09}.  In addition to this type of application, we demonstrate how harmonic descriptors can be applied to the broad range of structural characterization problems to which bond order parameters have traditionally been applied for simpler systems.  

This article is organized as follows.  In section~\ref{sec:hoppatterns}, we describe how to extract patterns from complex particle systems that can be described by harmonic shape descriptors.  In section~\ref{sec:hopdescriptors} we describe, in detail, how the descriptors can be computed mathematically, and explore their unique properties, such as rotational invariance and sensitivity to rotational symmetries.  In section~\ref{sec:hopapplications}, we describe how the harmonic descriptors can be used to solve representative problems from the field of computational and experimental self-assembly and computational biology.  In section~\ref{sec:hoptricks}, we explore how the special symmetry properties of harmonic descriptors can be applied to solve unique problems that are not easily solved by other types of shape descriptors.  The methods that we describe here are applicable to all types of particle systems, both simulated and experimental, for which particle positions are known or can be accurately imaged.  To aid with the dissemination of these techniques and new algorithms using harmonic shape descriptors as a basis, we provide a software library via the web\cite{smwebsite}.   A general review of the topic of applying shape matching algorithms to particle systems can be found in References~\cite{keys10-ARCMP}~and~\cite{keys10-LONG}.

\section{Pattern Extraction}
\label{sec:hoppatterns}


The first step in the structure characterization schemes that we will employ is to extract a representative structural pattern from the particle system that can be ``indexed'' (i.e., described mathematically) by a harmonic shape descriptor.  While this is relatively trivial for small clusters of spherical point particles\cite{snr83}, for complex structures, some physical and mathematical intuition is often required.  As we will demonstrate in detail in section~\ref{sec:hopdescriptors}, the harmonic shape descriptors that we introduce are best suited to index patterns on the unit circle, sphere, disk or ball.   Such representations are often sufficient to distinguish between even very similar structures with a high degree of precision.  The patterns may represent the particles themselves or some interesting pattern formed by their positions or density profile.  In some cases, the raw data may be preprocessed to better extract important structural features, for example, by spatial coarse-graining, time averaging, or potential energy minimization.

Structural patterns in general can be described by a set of positions $\{\textbf{x}\} = \{\textbf{x}_1,\textbf{x}_2, \dots \textbf{x}_n \}$ and corresponding weights $\{f\} = \{f_1,f_2, \dots f_n \}$.  For point cloud data, or raw particle coordinates, $\{\textbf{x}\}$ represents the particle positions and the weights $\{f\}$ are equivalent and usually taken to be 1.  For voxel data (i.e., volumetric data, often used to describe density maps), $\{\textbf{x}\}$ represents the positions of bins on a grid with weights $\{f\}$.  The same representation can be used to describe experimental images; in this case, $\{\textbf{x}\}$ represents the positions of the pixels and $\{f\}$ represents their intensity.  This notation allows us to write general equations for shape descriptors in section~\ref{sec:hopdescriptors}.

\subsection{Local Structures}

\begin{figure}
\begin{center}
\includegraphics[width=0.8\columnwidth]{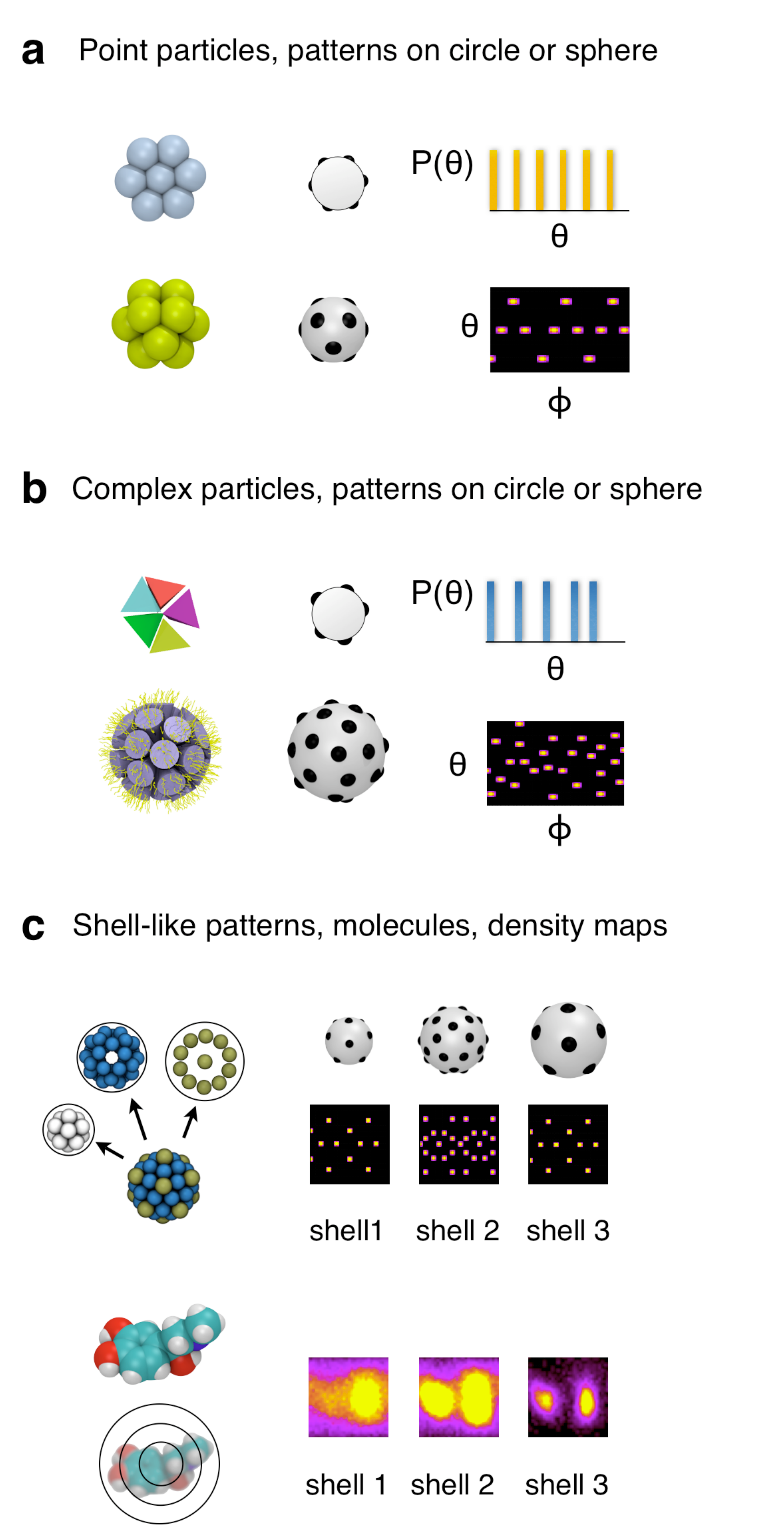}
\caption{Extracting local patterns. (\textit{a}) Clusters of point particles are expressed as patterns on the perimeter of the circle or on the surface of the sphere. (\textit{b})  More complex particles are often idealized as point particles to the same effect.  (\textit{b}) Point clusters with $r$-dependence or spatial density maps can be decomposed into independent patterns for multiple radial shells.}
\label{fig:hop2}
\end{center}
\end{figure}

As originally shown in the context of bond order parameters\cite{halperin78, snr83}, a cluster of point particles is one simple type of structure that can be trivially represented by the projection of the points onto the surface of a circle or sphere.  The patterns for two different point clusters are shown in Fig.~\ref{fig:hop2}a.  Notice that we exclude the center particle from the pattern, since it has no specific direction relative to itself.  The clusters need not strictly consist of point particles; so long as particle shape is not important, the structural pattern can be described by placing points at the particle centroids, as depicted in Fig.~\ref{fig:hop2}b.   Often, local structures are isolated from the bulk system by applying a clustering algorithm.  One standard scheme is to cluster all particles within a cutoff range\cite{snr83, iac07}.  More specialized schemes can be applied for specific applications, such as particles with complex shape\cite{amir09}.

Projecting patterns onto the circle or sphere neglects radial information.  Therefore, such projections can lead to non-distinguishing patterns for structures with radial dependence.  One solution is to decompose structures into concentric shells, projecting each shell onto the circle or sphere independently, as depicted in Fig.~\ref{fig:hop2}c.  Structures can be compared by matching corresponding shells.  Alternatively, complex patterns can be represented on the unit disk or unit ball.  These representations account for radial dependence as well as angular dependence.  As described in section~\ref{ssec:hopzernike}, representation on the disk or ball is optimal when decomposition into shells is inaccurate, or gives degenerate representations.

\subsection{Global Structures}

\begin{figure*}
\begin{center}
\includegraphics[width=0.7\textwidth]{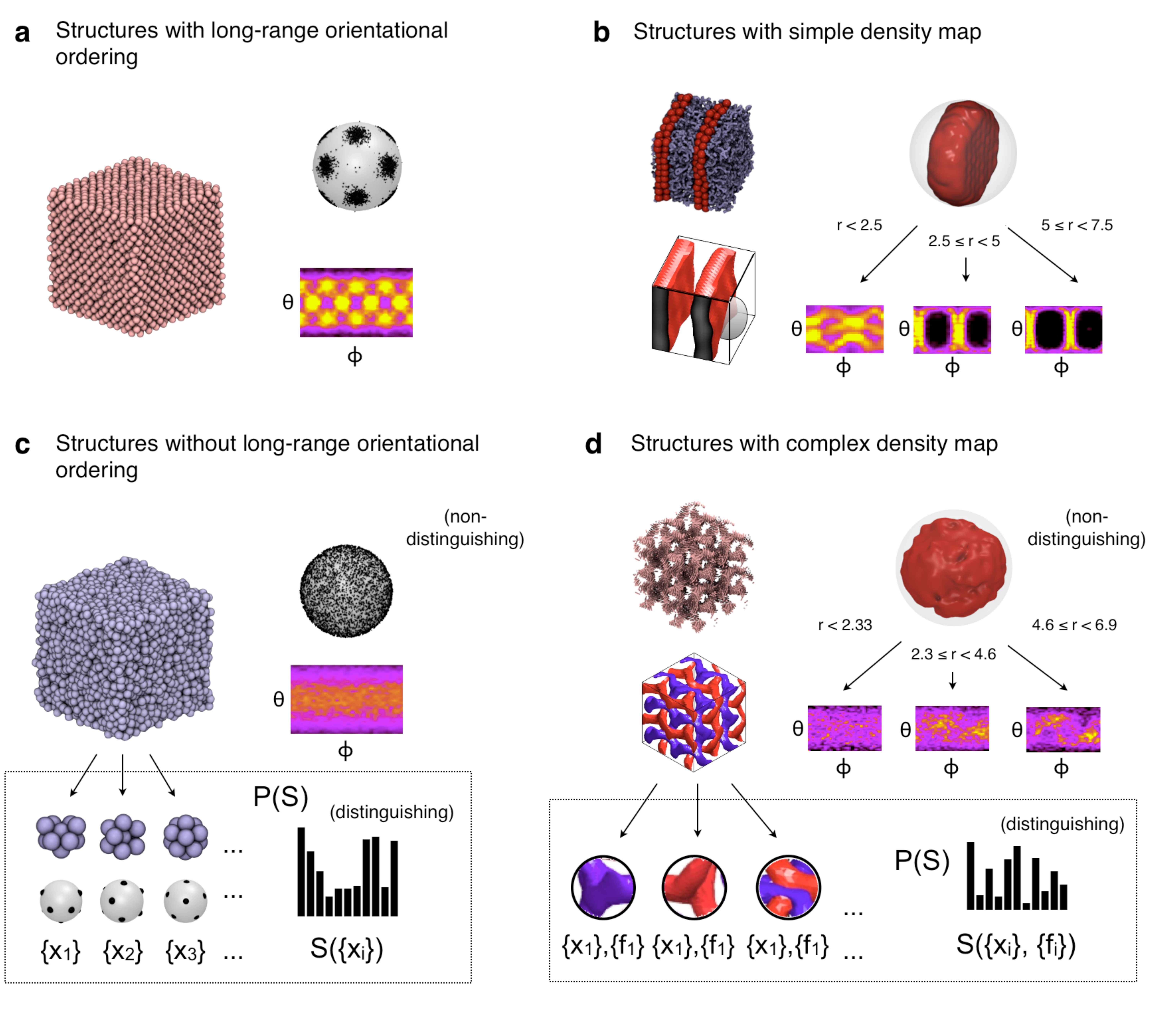}
\caption{Extracting global patterns. (\textit{a}) Superposition of local patterns for an fcc crystal.  (\textit{b}) Superposition of local patterns for a phase separated lamellar structure formed by tethered nanospheres\cite{iac07}. (\textit{c}) Superposition and probability distribution of local patterns for a liquid.  Since the structure has no long range orientational ordering, superposition is non-distinguishing.  (\textit{d})  Superposition and probability distribution of local patterns for a double-gyroid structure formed by tethered nano rods\cite{horsch2005, gyroid}.  For this complex structure, the superposition of local patterns is non distinguishing, even though the structure has long-range ordering.}
\label{fig:hop3}
\end{center}
\end{figure*}

Structures with long-range ordering, such as crystals and phase-separated structures, cannot be directly indexed on the circle, sphere, disk or ball.  Rather global patterns are constructed by combining different pieces of local information.  One such method, originally introduced in the context of bond order parameters\cite{snr83}, is based on computing the superposition of all local patterns in a sample\cite{roth2000}.  Fig.~\ref{fig:hop3}a shows the superposition of local patterns in a face centered cubic crystal, where local patterns are defined by the projection of neighbor directions on the unit sphere.  Since crystals have long-range orientational ordering, the neighbor directions are coincident throughout the sample.  This type of structural pattern is independent of the shape of the underlying particles, and thus can be applied to many assembled structures, such as crystals of patchy particles\cite{zhenlidiamond}, polyhedral particles\cite{amir09}, or phase-separated structures that form micelles or cylinders arranged in crystalline superlattices, such as dendrimers\cite{ungar03}, block copolymers\cite{yoon2005}, or tethered nanoparticles\cite{horsch2005, tnv, tns, ditethered, iac07}.  This superposition scheme is also applicable to many non-crystalline global phase-separated structures, such as layered and network structures formed by block copolymers or tethered nanoparticle systems\cite{horsch2005, tnv, tns}.  Fig.~\ref{fig:hop3}b shows the superposition of local density maps represented on the unit ball for a phase-separated sheet structure formed by tethered nano spheres\cite{tns}.

For structures without long-range orientational ordering, the superposition of local patterns results in a random pattern over long ranges.  Fig.~\ref{fig:hop3}c shows the superposition of local patterns for an atomic liquid, which results in a uniform distribution on the sphere.  Since the same pattern is inherent to all liquids, gases, gels, etc., regardless of the underlying particle shape, superposition gives no structural information, other than indicating an absence of long-range orientational order.  Orientationally-disordered structures of all types can be differentiated by considering the probability distribution functions of local patterns, rather than the superposition.  The probability distribution scheme is also useful when comparing structures with complex local patterns where superposition becomes degenerate or non-distinguishing, such as complex network structures\cite{gyroid}.  This is depicted for the double-gyroid structure formed by tethered nano-rods\cite{horsch2005, gyroid} in Fig.~\ref{fig:hop3}d.

\section{Harmonic Descriptors}
\label{sec:hopdescriptors}


Given a pattern on the unit circle, sphere, disk or ball, the harmonic shape descriptors reviewed in this section can be used to index the pattern into a compact vector representation.  This vector, or ``shape descriptor,'' can be compared with other shape descriptors to obtain a quantitative measure of similarity between structures.  In the derivation that follows, we introduce harmonic descriptors from a perspective that draws on elements of shape matching and signal processing.  This contrasts with the physics-based perspective presented in the original derivation of bond order parameters\cite{snr83}.  The alternate perspective is meant to highlight the fundamentally different way in which the structural metrics are used; whereas bond order parameters are often applied directly as order parameters, harmonic descriptors represent structural fingerprints that must be matched to obtain structural information, which can subsequently be used for constructing order parameters.

Before we introduce harmonic descriptors, it is important to understand why they are more useful than simpler shape descriptor methods.  Consider, for example, the problem of mathematically comparing two simple structures, such as the clusters shown in Fig.~\ref{fig:hop2}a.  Perhaps the most obvious way to describe the different structural patterns is to simply use the coordinates themselves as the shape descriptor.  While this greatly simplifies the initial step of creating a shape descriptor, it complicates matching significantly, since we do not typically know \textit{a priori} the optimal correspondence between the coordinates in different lists.  Reordering the lists is an optimization problem that can be solved, for example, by applying the Hungarian method\cite{hungarian}.  This type of problem scales as $O(N^3)$ and thus quickly becomes inefficient for large $N$.  A simple solution to the assignment problem is to create a probability histogram on the unit circle, sphere, disk or ball, as depicted in Figs.~\ref{fig:hop2}~and~\ref{fig:hop3}.  Since the histogram bins are independent of the order of the particles in the list (and number of particles), no assignment is required.  Although this representation, known as the ``shape histogram\cite{ankerst},'' is very useful for some applications, one drawback is that to compare patterns in a way that is rotation-invariant, the patterns (or the histograms) must be aligned prior to matching.  This ``registration\cite{icp, pca}'' step is computationally expensive and potentially inaccurate if applied naively.  One elegant solution to these problems is to compute the discrete Fourier transform (DFT) for each shell in the shape histogram\cite{ylm}.  The DFT transforms the pattern into its frequency-domain representation, which can be used to obtain rotation-independent harmonic descriptors through a simple mathematical operation as described in the following section.  As an additional advantage, harmonic descriptors have adjustable frequency parameters that can be tuned to highlight important rotational symmetries or give a variable degree of coarse-graining.  The manner in which we compute the harmonic descriptors depends on the coordinate system best used to describe the structure.  In the following sections, we first introduce Fourier descriptors\cite{fourier, ylm, ylm2}, which are suited for indexing shapes on the unit circle ($\theta$ dependence) or sphere ($\theta,\phi$ dependence), and then introduce Zernike moments which are suited for indexing shapes on the unit disk ($r, \theta$ dependence) or ball ($r, \theta, \phi$ dependence).


\subsection{Fourier Descriptors}

\begin{figure}
\begin{center}
\includegraphics[width=0.9\columnwidth]{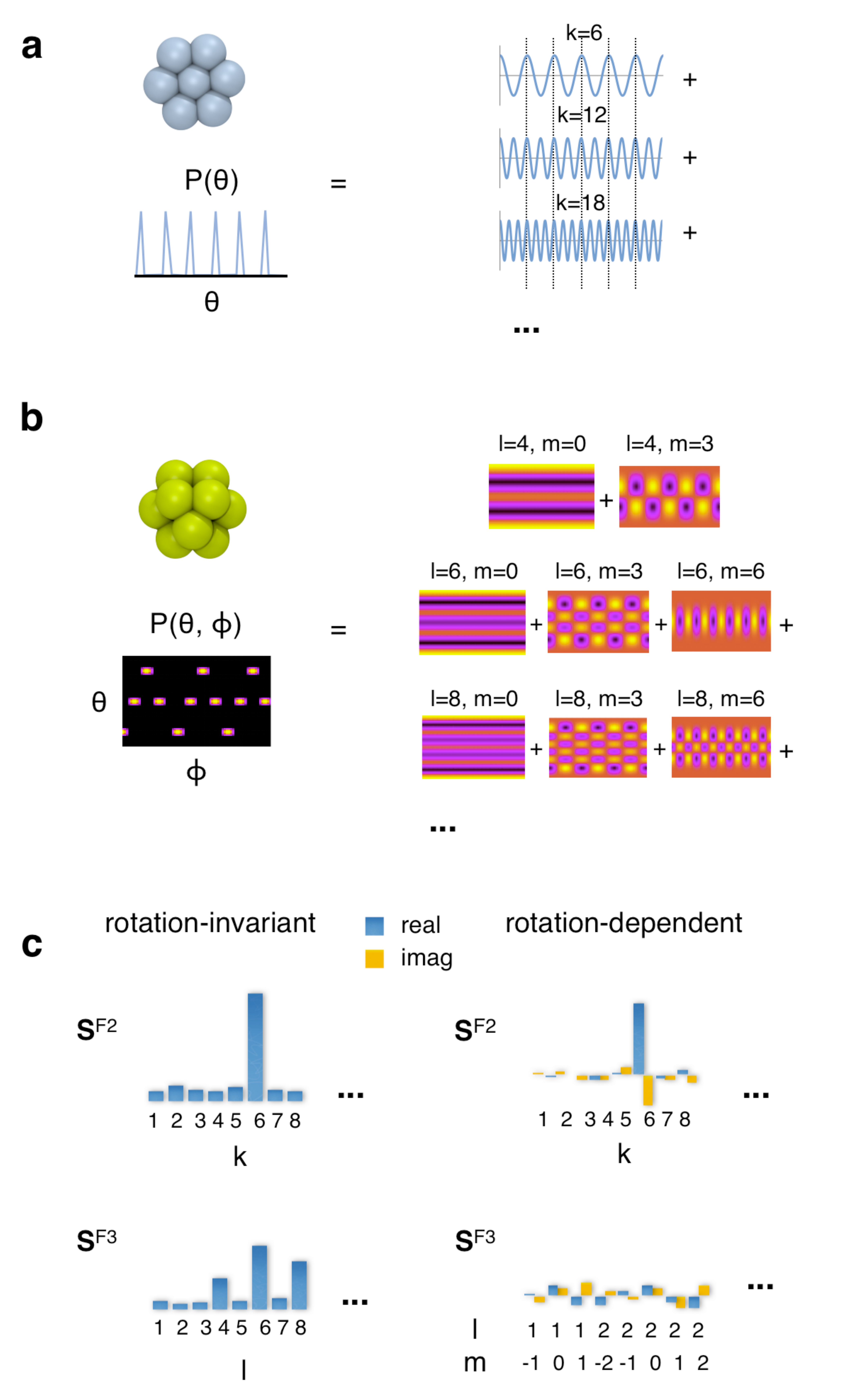}
\caption{Fourier descriptors. (\textit{a}) Decomposition of the angular pattern obtained for a 2d hexagonal cluster into a sum of sines and cosines.  Since the cluster has 6-fold symmetry, the leading coefficients are given by multiples of $\ell=6$. (\textit{b}) Decomposition of the angular pattern obtained for a 3d fcc cluster into a sum of spherical harmonics.  Since the cluster has 4 and 6-fold symmetry, the leading coefficients are given by multiples of $\ell=4$ and $\ell=6$.  (\textit{c}) Schematic of the resulting Fourier descriptors.  The value of the each component in the descriptor is proportional to the contribution of the corresponding harmonic to the overall sum. Notice that rotation-invariant Fourier descriptors contain real, positive components, while rotation-dependent descriptors contain complex components.}
\label{fig:hop4}
\end{center}
\end{figure}

Fourier descriptors are designed to efficiently index structural patterns on the circle or sphere.  A Fourier transform represents a function as a sum of harmonic components. For a pattern along the 1d perimeter of the circle, this representation is performed by summing complex exponential terms:
\begin{equation}
f(\theta_j)= \sum_{\ell=0}^{\ell_{max}} \psi_\ell \exp{\left[ i \ell \theta_j \right]} \quad j=1, 2,... n_{bin}.
\end{equation}
Here, $f(\theta_j)$ is the intensity of the pattern at a particular point along the perimeter of the circle $\theta_j$.  The terms $\psi_\ell$, known as ``Fourier coefficients,'' indicate the strength of the pattern for a particular frequency $\ell$. Typically, we cut off the frequency $\ell$ at some finite value $\ell_{max}$, since the information for high-frequency $\ell$ becomes increasingly dominated by noise in the structure.  The $\ell = 0$ and $\ell = 1$ terms only contain information regarding the position and center of mass of the pattern and are sometimes excluded.  If the pattern consists of more than one radial shell, we compute the Fourier transform for each shell independently.

The Fourier coefficients contain structural information that can be used to create shape descriptors with various properties.  The coefficients are given by:
\begin{equation}
\label{eq:hoppsi}
\psi_\ell = \frac{1}{n_{pts}} \sum_{j=1}^{n_{pts}} f_j \exp{\left[i \ell \theta_j\right]^*}.
\end{equation}
The term $\theta_j$ is the angle of an input point $\textbf{x}_j$ with intensity $f_j$ from our pattern $\{\textbf{x}\}, \{f\}$.  The coefficients $\psi_\ell$ are complex numbers.  The $^*$ denotes the complex conjugate.  As outlined in the previous section, the input points from our pattern can represent either the positions of bins in the circular shape histogram or the raw data if no binning is performed.  The representations become equivalent as the bin size approaches zero and only one point can occupy a given bin.

Since the Fourier transform is a frequency-domain representation of the pattern, it gives the strongest signal for frequencies that reflect periodicities in the pattern around the circle.  That is, patterns with $n$-fold rotational symmetry yield high values of $\psi_\ell$ (and, as we will discuss later, $\textbf{q}_\ell$).  Since rotational symmetries are relatively insensitive to small changes to the pattern, Fourier descriptors are relatively insensitive to thermal noise, particularly for low-frequency coefficients.  Although the Fourier coefficients in their complex number form are not rotation-invariant, we can convert them to their invariant form by computing the magnitude of each coefficient.  The invariant circular coefficients are given by:
\begin{equation}
|\psi_\ell| = \psi_\ell \psi_\ell^* = \sqrt{\Re(\psi_\ell)^2 + \Im(\psi_\ell)^2}.
\end{equation}
The Fourier invariants are positive real numbers.  

\begin{figure}
\begin{center}
\includegraphics[width=1.0\columnwidth]{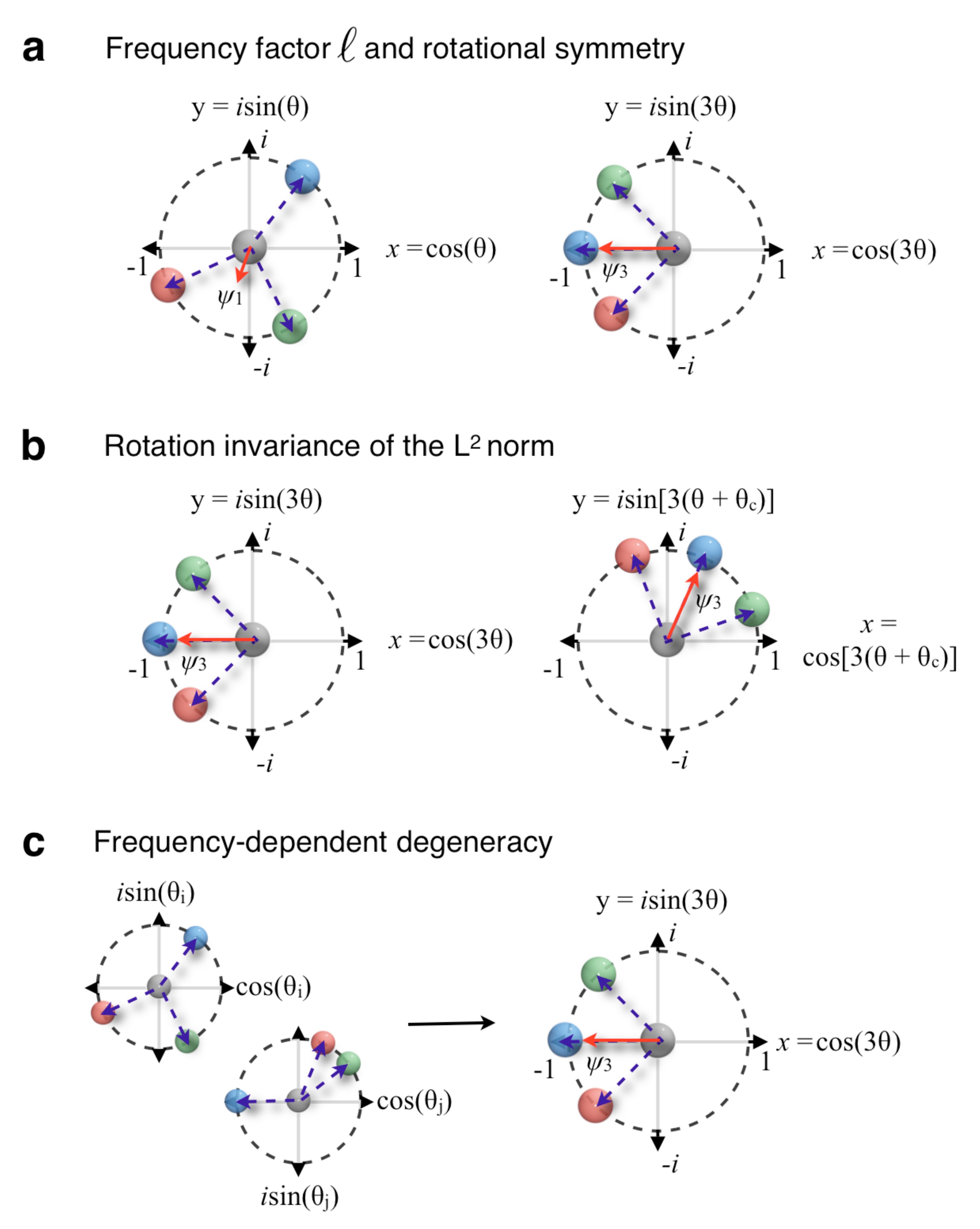}
\caption{Properties of harmonic descriptors. (\textit{a}) The depicted cluster of four particles is somewhat  3-fold symmetric.  When each angle $\theta$ is multiplied by a frequency factor $\ell=3$, the centroid position moves farther from the origin ($| \psi_3| > | \psi_1| $), as depicted by the red vectors.  (\textit{b})  When the cluster is rotated (i.e., a constant factor $\theta_c$ is added to each angle $\theta$), $| \psi_3|$ remains constant, although $\psi_3$ changes. (\textit{c}) When the clusters on the left are scaled by a frequency factor $\ell=3$, the transformed centroids $\psi_3$ become equivalent, even though the underlying shapes are different.}
\label{fig:hop5}
\end{center}
\end{figure}

To obtain some physical understanding of the properties of Fourier coefficients, consider the small cluster in Fig.~\ref{fig:hop5}a.  If we normalize the cluster to the unit circle, the centroid (unweighted center of mass) is given in complex coordinates by the conjugate of the $\ell=1$ coefficient $\psi_1^*$.  Using a frequency term other than $\ell=1$ multiplies each angle $\theta$ by a factor, effectively stretching or compressing the pattern along the circle.  We see that choosing $\ell = \psi_n$, where $\psi_n$ is a rotational symmetry of the cluster, causes the different angles $\theta_j$ to coincide, resulting in a non-vanising centroid for the transformed cluster.  Thus, the Fourier coefficient with $\ell \neq 1$ represents the centroid of the stretched or compressed pattern. Although the position of the centroid is dependent on the cluster orientation, this distance from the origin to the centroid is invariant under rotations.  Thus, we obtain a rotation-invariant descriptor by computing the magnitude of the coefficient.

These properties of Fourier coeffcients have been exploited in the context of bond order parameters\cite{halperin78, snr83}.  For example, particular coefficient magnitudes, such as $\psi_6$ (or, analogously $\textbf{q}_4$ and $\textbf{q}_6$, see below) have been used directly as scalar order parameters\cite{snr83,gubbins}.  This method is often sufficient for the simple structures that we encounter in systems of spherical particles.  However, more complex structures often require a full range of coefficients, and order parameters must be constructed by comparing sets of coefficients, rather than evaluating particular coefficients.  To create a descriptor from the Fourier coefficients, we simply combine the desired $\psi_\ell$ or $|\psi_\ell|$ into a long vector.  For example, a general rotation-invariant shape descriptor that is applicable to patterns on the circle over a range of symmetries is given by:
\begin{equation}
\textbf{S}^{\textrm{F2}} = <|\psi_0|, |\psi_1|, ... |\psi_{ \ell_{max} } |>.
\end{equation}
It is easy to imagine how different combinations of the Fourier descriptors can be used to create shape descriptors with different levels of robustness and sensitivity to particular symmetries.  For many applications it is common to include only coefficients with specific symmetries, or to use rotation-dependent coefficients.  

As outlined in the previous section, many 3d structures are well represented by patterns on the surface of the sphere.  The analogy to the 1d DFT on the 2d surface of the sphere is known as the discrete spherical harmonics transform (DSHT)\cite{fsht}, given by:
\begin{eqnarray}
f_j(\theta_j, \phi_j) =  \sum_{\ell=1}^{\ell_{max}} \sum_{m=-\ell}^{\ell} q_\ell^m  Y_{\ell}^m(\theta_j, \phi_j). \nonumber \\
\quad j=1, 2,... n_{bin}
\end{eqnarray}
The terms $Y_{\ell}^m[\theta_j, \phi_j]$ are spherical harmonics, defined by $Y_{\ell}^m(\theta, \phi) = N_\ell^m P_\ell^m (\cos \theta) \exp{(i m \phi)}$.  The term $N_\ell^m$ is a normalization factor $\sqrt{(2\ell+1) (\ell-m)! / (\ell+m)!} $ and $P_\ell^m$ is a Legendre polynomial.  We see that the DSHT is similar to the DFT except for an additional term depending on the polar angle $\theta$.

The Fourier coefficients for the DSHT are given by:
\begin{eqnarray}
\label{eq:hopq}
\textbf{q}_\ell =  \frac{1}{n_{pts}} \sum_{j=1}^{n_{pts}}  f_j N_\ell^m Y_\ell^{m*} (\theta_j, \phi_j). \nonumber \\
\quad m=-\ell, -\ell+1, ...\ell
\end{eqnarray}
Unlike the circular coefficients $\psi_\ell$, which are complex numbers, the spherical coefficients $\textbf{q}_\ell$ are $2\ell+1$ dimensional complex vectors.  Like the circular coefficients, the spherical Fourier coefficients are robust under noise, sensitive to rotational symmetries corresponding to $\ell$, and can be used to construct invariants.  Although the geometrical interpretation of these properties is more complex than for the 1d case, the same principles apply.

The rotation-invariant version of the spherical coefficients is given by:
\begin{equation}
|\textbf{q}_\ell| =  \sqrt { \frac{4\pi}{2\ell+1}\sum_{m=-\ell}^\ell |q_\ell^m|^2 }.
\end{equation}
Like the circular invariants $|\psi_\ell|$, the spherical invariants $|\textbf{q}_\ell|$ are positive real numbers.

In analogy with our example above, we can create a rotation-invariant shape descriptor for a pattern on the sphere by:
\begin{equation}
\textbf{S}^{\textrm{F3}} = <|\textbf{q}_0|, |\textbf{q}_1|, ... |\textbf{q}_{\ell_{max}}|>.
\end{equation}
Again, the optimal descriptor for a particular application depends on the desired properties such as robustness, sensitivity to specific symmetries, and rotation invariance.

Notice that although we use the notation ``$\psi_\ell$'' and ``$\textbf{q}_\ell$'' to highlight the connection with bond order parameters, we have redefined the order parameters slightly, by changing the sign of the complex exponential (i.e., the conjugates in equations~\ref{eq:hoppsi} and~\ref{eq:hopq}).  This sign change is inconsequential to the properties of the coefficients; our redefinition simply allows us to highlight the important relationship between $\psi_\ell$ and $\textbf{q}_\ell$ and the DFT, which is standard and extensively studied\cite{fft}.  An overview of the Fourier descriptors method is given in Fig.~\ref{fig:hop5}.  As a general rule of thumb, ``$\psi_\ell$'' is used when we only wish to describe the 2d ordering of a system (e.g., the spatial ordering within a single plane in a confined fluid\cite{gubbins, confinedfluids} or the crystalline arrangement of cylindrical domain\cite{ditethered, iacovella2009b}) and ``$\textbf{q}_\ell$'' is used when we wish to describe the 3d ordering (e.g., the spatial ordering in a 3d crystal\cite{tenwolde96} or structure of a compact 3d cluster\cite{iac07}).


\subsection{Zernike Descriptors}
\label{ssec:hopzernike}

As mentioned in the previous section, when patterns cannot be properly represented by the surface of a single circle or sphere, one solution is to break up the pattern into independent radial shells, and compute the Fourier descriptor for each shell independently.  We then construct a shape descriptor by combining the Fourier descriptors for each shell into a long vector:
\begin{equation}
\textbf{S}^{\textrm{F, multishell}} =  < \textbf{S}^F_{shell_1}, \textbf{S}^F_{shell_2}, ... \textbf{S}^F_{shell_n} >.
\end{equation}
Here, $\textbf{S}^\textrm{F}$ represents a Fourier descriptor, either $\textbf{S}^\textrm{F2}$ or $\textbf{S}^\textrm{F3}$, as defined in the previous section.

\begin{figure}
\begin{center}
\includegraphics[width=0.7\columnwidth]{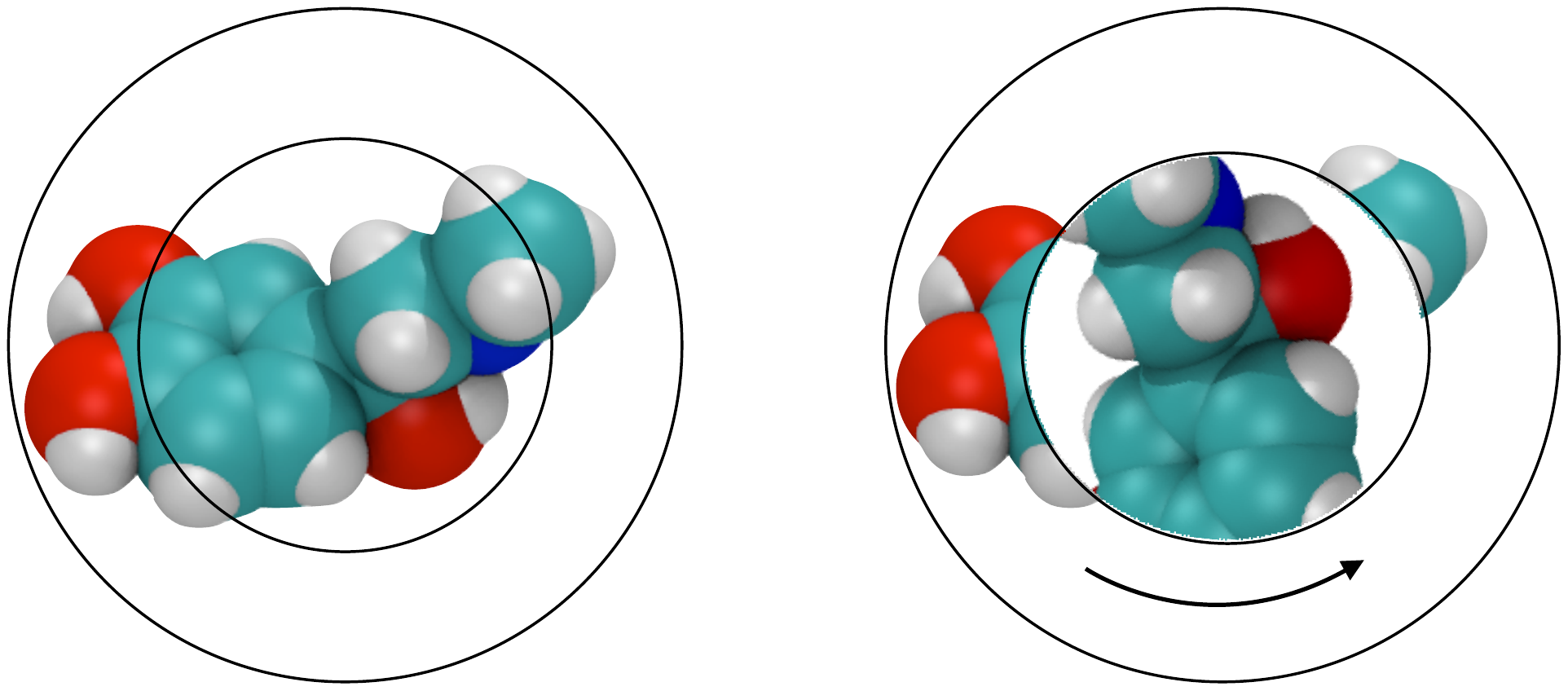}
\caption{Drawbacks of Fourier descriptors. Rotation-invariant Fourier descriptors for structures with multiple shells are insensitive to the relative orientation of inner and outer shells.  Using invariant Fourier descriptors to describe the two structures shown in the schematic shown erroneously produces identical results.}
\label{fig:hop6}
\end{center}
\end{figure}

While this scheme is sufficient for many problems, it has the drawback that small perturbations to the particle positions can cause maxima in the pattern to shift between shells, causing errors in matching, particularly when $n_{pts}$ is small.  Another drawback is that, since the shells are treated independently, the rotation-invariant Fourier descriptors are insensitive to relative orientations between the different shells; this is depicted in Fig.~\ref{fig:hop6} for two structures that would erroneously have matching descriptors.

We can solve these problems by representing our pattern in a coordinate system with radial dependence, which allows us to properly index patterns defined on the disk or ball rather than the unit circle or sphere.  To do so, we use Zernike radial polynomials in our expansion\cite{zernike2d, zernike3d}.  

We can express the intensity of a pattern at a given point on the unit disk as:
\begin{eqnarray}
f(r_j, \theta_j) = \sum_{n}\sum_{\ell}  a_{n\ell}  R_{n \ell}(r_j) \exp{\left[ i \ell \theta_j \right]}.  \nonumber \\
\quad j=1, 2, \dots n_{bin}
\end{eqnarray}
The terms $\theta_j$ and $r_j$ represent the position of a point on the unit disk.  The value of $\ell$ is restricted such that $\ell \leq n$ and $(n - \ell )$ is an even number.  The expansion coefficients $a_{n\ell}$ are known as ``Zernike moments'' and can be considered analogous to Fourier coefficients for the $r,\theta$ coordinate system.  The function $R_{n \ell}(r)$ is a radial polynomial, where $r$ is the radial distance from the center of the disk.  Thus, the 2d Zernike expansion is very similar to the 1d Fourier expansion, but with an additional radial term.

The Zernike moments are given by\cite{zernike2d}:
\begin{equation}
a_{n\ell} = \frac{n+1}{n_{pts} \pi} \sum_{j=1}^{n_{pts}} f_j(r_j, \theta_j) R_{n \ell}(r_j) \exp{\left[- i \ell \theta_j\right]}.
\end{equation}
The terms $\theta_j$ and $r_j$ represent the position of an input point $\textbf{x}_j$ in polar coordinates, normalized on the unit disk.  Again, we require that $\ell \leq n$ and $(n - \ell )$ is even.  Each moment is a complex number.
Since the radial polynomial is only dependent on $r$, the same invariance relations hold for the Zernike moments as for the Fourier coefficients.  To define a rotation-invariant moment on the disk, we take the complex magnitude of the moment:
\begin{equation}
|a_{n\ell}| = a_{n\ell} a_{n\ell}^*.
\end{equation}
The 2d Zernike invariants are positive real numbers.  We can create a Zernike descriptor by combining many moments in a vector.  For example, we can create a 2d rotation-invariant Zernike descriptor by:
\begin{equation}
\textbf{S}^{Z2} = <|a_{00}|, |a_{11}|, |a_{20}|, |a_{22}|, ... |a_{\ell_{max} \ell_{max}}|>.
\end{equation}
Like the Fourier descriptors, the frequency parameter $\ell$ has a straightforward relationship with the rotational symmetry of the pattern.  Thus, we can sometimes choose important moments \textit{a priori}.  However, we typically take all moments with $\ell$ within a limiting frequency $\ell_{max}$.

We can express a pattern on the unit ball as the sum of 3d Zernike moments:
\begin{eqnarray}
f(r_j, \theta_j, \phi_j) = \sum_n \sum_\ell \sum_m z_{n\ell}^m R_{n \ell}(r_j) Y_\ell^m (\theta_j, \phi_j). \nonumber \\
\quad j=1, 2, ... n_{bin} 
\end{eqnarray}
The terms $[r_j, \theta_j, \phi_j]$ give the position of a point on the unit sphere.  Again, we require $\ell \leq n$ and $(n - \ell )$ is even.  The moments are defined similarly to the Fourier coefficients on the surface of the sphere, but again with an additional radial component.
The 3d Zernike moments are given by\cite{zernike3d}:
\begin{eqnarray}
\textbf{z}_{n\ell} = \frac{3(n+1)}{4 n_{pts} \pi} \sum_{j=1}^{n_{pts}} f_j R_{n \ell}(r_j) N_\ell^m Y_\ell^{m*} (\theta_j, \phi_j). \nonumber \\
\quad m=-\ell, -\ell+1, ...\ell
\end{eqnarray}
The variables $[r_j, \theta_j, \phi_j]$ represent the position of an input point $\textbf{x}_j$ in spherical coordinates, normalized on the unit sphere.  Whereas the 2d Zernike moments $a_{n\ell}$ are  complex numbers and the 3d Zernike moments $\textbf{z}_{n\ell}$ are complex vectors of length $2\ell + 1$.
Analogously to the spherical Fourier coefficients, we take the magnitude of the complex vector $|\textbf{z}_{n\ell}|$ to define invariant moments on the unit ball:
\begin{equation}
|\textbf{z}_{n\ell}| =  \sqrt { \frac{4\pi}{2\ell+1}\sum_{m=-\ell}^\ell |z_{n\ell}^m|^2 }.
\end{equation}
The 3d Zernike invariants are positive real numbers.  Like the Fourier coefficients, they are sensitive to the rotational symmetries of the pattern and are robust under small perturbations.  We can create a rotation invariant symmetry-independent 3d Zernike descriptor according to:
\begin{equation}
\textbf{S}^{Z3} = <|\textbf{z}_{00}|, |\textbf{z}_{11}|, |\textbf{z}_{20}|, |\textbf{z}_{22}|, ... |\textbf{z}_{\ell_{max} \ell_{max}}|>.
\end{equation}

\begin{figure}
\begin{center}
\includegraphics[width=0.9\columnwidth]{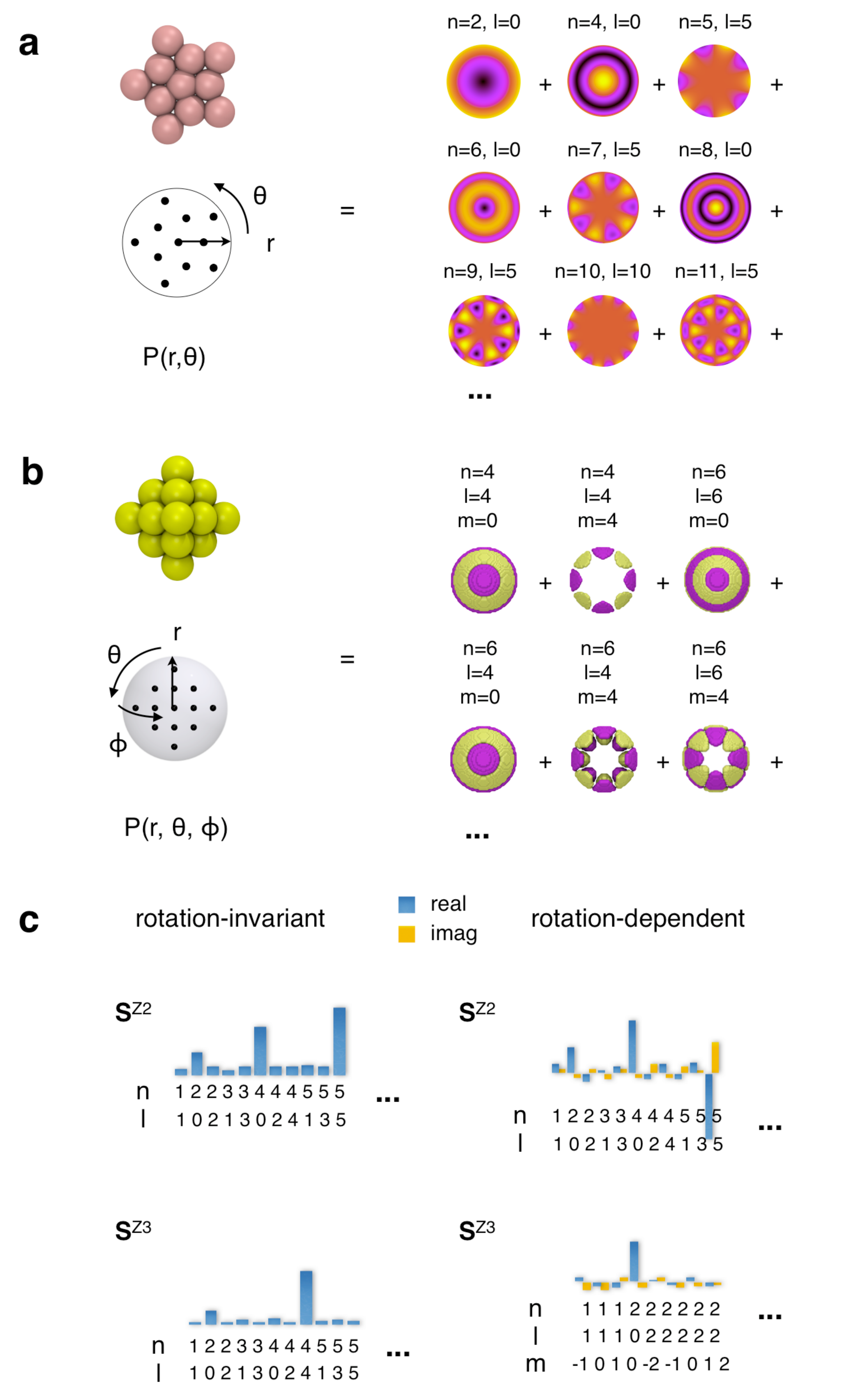}
\caption{Zernike descriptors.  (\textit{a})  Decomposition of the pattern obtained for a 2d pentagonal cluster with Zernike moments.  Since the cluster has 5-fold symmetry, the leading coefficients are given by multiples of $\ell=6$.  (\textit{b})  Decomposition of the pattern obtained for a 3d fcc cluster with Zernike moments.  Since the cluster has 4 and 6-fold symmetry, the leading coefficients are given by multiples of $\ell=4$ and $\ell=6$.  (\textit{c}) Schematic of the resulting Zernike descriptors.  The value of the each component in the descriptor is proportional to the contribution of the corresponding moment to the overall sum. Notice that rotation-invariant Zernike descriptors contain real, positive components, while rotation-dependent descriptors contain complex components.}
\label{fig:hop7}
\end{center}
\end{figure}

When computing either multiple-shell Fourier descriptors or Zernike moments it is essential that the patterns being compared are normalized consistently.  In the case of Zernike moments, all points in $\{ x \}$ must lie on the unit ball or disk.  Typically, patterns are normalized by translating the centroid to the origin and rescaling the coordinates such that every point on the pattern has a radial distance less than 1.  This scheme is sufficient for the majority of patterns that we encounter in particle systems.  An overview of the Zernike scheme is depicted in Fig.~\ref{fig:hop7}.

\subsection{Computational Considerations}
Fourier and Zernike coefficients can be computed from either point cloud data (i.e., raw particle positions) or voxel data (i.e., volumetric data or pixel data).   As mentioned previously, the two representations are essentially equivalent; point cloud data represent the limit of zero bin size, where each $f_i$ is equivalently 1.  While this distinction does not affect the properties or definition of the descriptors, it becomes important when considering the computational cost of a matching application.  If the input data is point cloud data, we must compute the descriptors for each structure independently.  However, for volumetric data, we can compute the contribution to each coefficient for each point on the grid beforehand, and then simply multiply by the intensity of each shape $f_i$ to compute the value of the coefficients.  This can greatly reduce the computational cost when $n_{pts}$ is large or many coefficients are used.  

As an additional consideration, the time required for computing the transforms themselves can be greatly reduced by computing the fast Fourier transform (FFT) rather than the DFT (or the equivalent for the appropriate coordinate system).  Methods for computing the FFT and the discrete spherical harmonics transform, respectively, are given in references~\cite{fft} and~\cite{fsht}.  An efficient method for computing Zernike coefficients is given in reference~\cite{zernike3d}.  

\section{Quantifying Similarity}

The shape descriptors derived in the previous section can be considered compact mathematical representations of the underlying particle structures.  The physical similarity between different particle structures can then be quantified by the mathematical similarity between shape descriptors.  Shape descriptor similarity is quantified by a similarity metric ``$M$,'' that gives a scalar value that is proportional to the similarity between descriptor pairs.  For convenience, we define $M$ such that, by construction, it lies on the interval $M \in [0,1]$, or sometimes $M\in[-1,1]$.  This strengthens the analogy between $M$ and what we normally consider to be an order parameter, since order parameters typically give a value of $1$ for perfectly ordered structures and $0$ for perfectly disordered structures.  Since our harmonic shape descriptors are defined as vectors, we can define similarity metrics based on standard vector operations.  For example, one simple similarity metric is given by the Euclidean distance between shape descriptor vectors:
\begin{equation}
M_{dist}(\textbf{S}_1, \textbf{S}_2) = 1 - | \textbf{S}_{1} - \textbf{S}_{2}| / (|\textbf{S}_1|  + |\textbf{S}_2|).
\end{equation}
Variations of $M_{dist}$ are common throughout the shape matching literature.  Another simple similarity metric is proportional to the dot product between shape descriptors: 
\begin{equation}
M_{dot}(\textbf{S}_1, \textbf{S}_2) = \textbf{S}_{1} \cdot \textbf{S}_{2} / ( |\textbf{S}_1| |\textbf{S}_2|).
\end{equation}
Schemes similar to $M_{dot}$ have been used in applications involving standard bond order parameters, such as measuring correlation lengths\cite{snr83, marcus96} and identifying crystal grains\cite{tenwolde96}.

Similar information is obtained from $M_{dist}$ and $M_{dot}$;  the only difference is that whereas $M_{dist}$ is more sensitive to the absolute difference between vector components, $M_{dot}$ is more sensitive to the overall direction of the vector and the sign of the components.  Thus, $M_{dot}$ is often superior when matching non-ideal structures from a particle system to mathematically perfect reference structures, since thermal noise will tend to damp the frequency domain signal, but the descriptor will retain the same basic character for a given class of structures and hence the same direction.  The $M_{dot}$ metric may also be favorable when comparing rotation-dependent harmonic descriptors, which may contain either negative or positive components, whereas $M_{dist}$ may be favorable for invariant descriptors, where all values are inherently positive.  Since compared shapes are usually at least grossly similar, matching values are rarely $0$ for either metric.  As is discussed in greater detail in reference~\cite{keys10-LONG}, it is often necessary to determine a lower bound on $M$ by comparing to structures that are known to match poorly to obtain a baseline value.

\section{Example Applications}
\label{sec:hopapplications}


Shape similarity information obtained from evaluating the match, $M$, between shape descriptors, can be applied to create various types of structural metrics for complex particle systems.  In this section, we provide several example applications that can be addressed by the use of shape matching with harmonic descriptors.  We provide a more extensive range of example applications based on shape matching methods in general in reference~\cite{keys10-LONG}.  Several additional examples of shape matching applications based on the $\textbf{F}^3$ descriptor are given in reference~\cite{keys10-ARCMP}.  

\subsection{Order Parameters and Correlation Functions}
\label{ssec:hopopcf}

\begin{figure}
\begin{center}
\includegraphics[width=0.9\columnwidth]{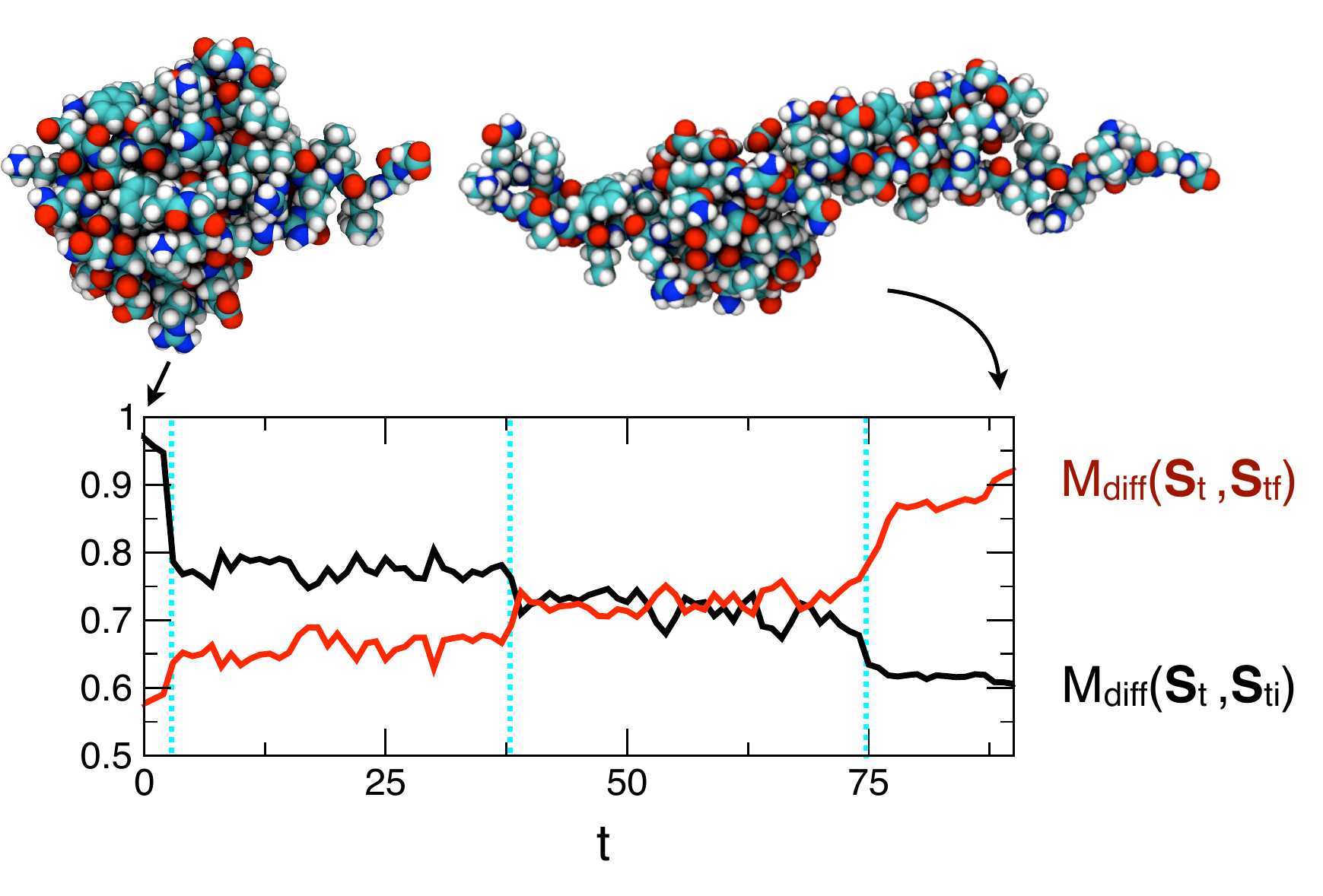}
\caption{Tracking structural transitions in the protein Ubiquitin as it is pulled from both ends\cite{namd, vmd}.  The protein is pulled linearly as a function of time.  The black curve shows a matching order parameter using the initial folded configuration $i$ as a reference structure.  The red curve shows the order parameter using the final unfolded configuration $f$ as the reference structure.   The blue dashed lines highlight the times at which significant structural changes occur.}
\label{fig:hop8}
\end{center}
\end{figure}

Perhaps the most standard application of order parameters is to track structural transitions,  either as a function of time or a changing reaction coordinate.   As an example, consider the protein ``Ubiquitin\cite{ubiquitin}'' shown in Fig.~\ref{fig:hop8}, which unfolds as it is pulled from both ends.  This is a standard example problem from the NAMD and VMD tutorials\cite{namd, vmd}, both of which are available online.  Since the structure is 3-dimensional and has radial dependence, we can index it using a Zernike descriptor on the unit ball, $\textbf{S}^\mathrm{Z3}$.  To match the shape independently of the orientation of the sheet, we take rotation invariant moments with $\ell$ in the range $4 \leq \ell \leq12$.  We use both the initial folded state $i$ and the final unfolded state $f$ as reference states.  Fig.~\ref{fig:hop8} shows the unfolding transition as a function of time $t$ in a NAMD molecular dynamics simulation.  The protein unfolds in three steps (blue dashed lines), in agreement with visual inspection.  The noise in the data is indicative of the thermal fluctuations in shape observed at this temperature. 

\begin{figure}
\begin{center}
\includegraphics[width=0.9\columnwidth]{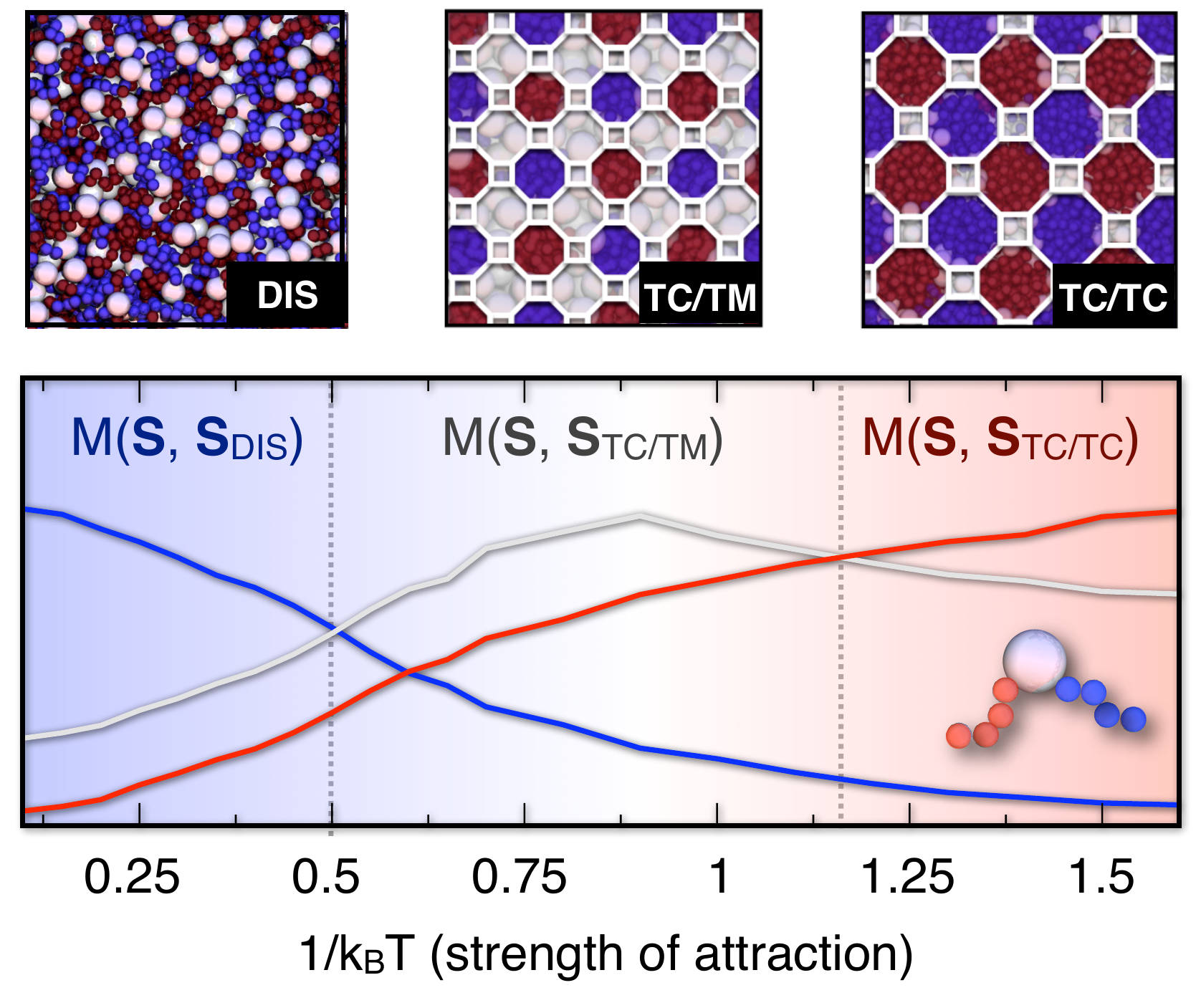}
\caption{Transitions in ditethered nanospheres. (\textit{a}) Depiction of the three structures formed by a ditethered nanosphere system\cite{iacovella2009b} (left to right: disordered, TC/TM, TC/TC). Notice that the cylindrical structures span the simulation box in the z-dimension, into the page.  (\textit{b}) Matching order parameter for the three reference structures as a function of inverse temperature.}
\label{fig:hop9}
\end{center}
\end{figure}

As a slightly more complex example of characterizing transitions, consider the micro-phase separated structures formed by ditethered nanospheres\cite{iacovella2009b} shown in Fig.~\ref{fig:hop9}.  The structure of the equilibrium system goes through two transitions as a function of the effective inverse temperature, first from a disordered structure to a tetragonal cylinder/tetragonal-mesh (TC/TM) phase and then to a similar tetragonal cylinder (TC/TC) phase. The abbreviations indicate the structure of the tethers (blue, red) and nanoparticles (white), respectively.  We can quantify this behavior by matching the global patterns obtained at different temperatures with ideal structures taken from within the three structural regimes: the disordered regime, the TC/TM regime, and the TC/TC regime.  The global pattern for each structure is characterized by the probability distribution of local density maps for each particle type, as depicted in Fig. ~\ref{fig:hop3}d.  To capture ordering on a range of length scales, density maps are computed for four radial shells with $r = [3\sigma, 4.5\sigma, \dots 9\sigma]$, where $\sigma$ is the characteristic lengthscale, given by the diameter of the tether beads in the simulation.  For each shell, we compute the rotation-invariant Fourier descriptors $\textbf{S}^{\textrm{F3}}$, where we take a range of frequencies $4 \leq \ell \leq 12$.  A pseudo-order parameter for each reference structure is then given by $M_{diff}(\textbf{S}^{\textrm{F3}}, \textbf{S}^{\textrm{F3}}_{ref})$.  Fig.~\ref{fig:hop8} shows the order parameters for the three reference structures as a function of inverse temperature.  We observe that the structural transition between the three phases is smooth and continuous, as verified by visual inspection.  In reference~\cite{keys10-LONG}, we show that, for this particular problem, we can obtain a nearly identical result using a simpler shape descriptor akin to the radial distribution function $g(r)$.  However, for more complex phase separated structures, harmonic descriptors typically give a better representation of the underlying shapes than such coarse measures as $g(r)$.

In addition to characterizing how structures change as a function of time or a reaction coordinate, another common application is to characterize how structures change in space by computing correlation functions.   In this case, we choose structures from different points in space, rather than ideal structures, as references.  Several examples of spatial correlation functions based on bond order parameters have already been defined in the context of measuring lengthscales for crystal-like ordering\cite{halperin78, nelsonc6, snr83, ernstnagelgrest}.   This involves measuring quantities such as $\left< M(\textbf{S}_i,\textbf{S}_j) \right>(r_{i,j})$, that give the average similarity value as a function of the radial separation $r_{i,j}$.  Typically, a particular rotation-dependent Fourier coefficient, for example $\psi_6$ or $\textbf{q}_6$, is chosen for $\textbf{S}$ and $M_{dot}$ is chosen for $M$.  An alternative, and well known, spatial correlation function based on bond order parameters is given by the $q_6 \cdot q_6$ crystal grain detection scheme of reference~\cite{tenwolde96}, which has been applied to studying nucleation and growth and characterizing crystalline defects in several simulation and experimental studies involving close-packed and bcc crystals\cite{auer04, gasser01, laura, tesfuv2}.  Here, crystal grains are identified within a bulk liquid by first noticing that, for many crystals, local clusters within crystal grains match with their neighbors in terms of \textit{both} shape and orientation, whereas clusters in the liquid do not.  Thus, pairs of particles $i,j$ in grains typically satisfy $M(\textbf{S}_i,\textbf{S}_j) > M_{cut}$, where $\textbf{S}$ is a rotation-\textit{dependent} harmonic descriptor and $M_{cut}$ is a sufficiently high value so as to exclude poor matches.  Even in the liquid, some pairs of particles inevitably satisfy $M(\textbf{S}_i,\textbf{S}_j) > M_{cut}$ due to random fluctuations.  Thus, a local indicator of crystal-like ordering is given by:
\begin{equation}
f_i = \frac{1}{n_{nbr}} \sum_j^{n_{nbr}} \Theta[M(\textbf{S}_i,\textbf{S}_j) - M_{cut}].
\end{equation}
Here, $\Theta$ is the Heaviside function.  Typically, we enforce $f_i \geq f_{cut}$, where the value of $f_{cut}$ is chosen to distinguish liquid and crystal-like configurations\cite{tenwolde96, auer04}.  Although the original scheme is based on the $\textbf{q}_6$ Fourier coefficient as a shape descriptor (for identifying fcc, hcp, and bcc crystals) and the $M_{dot}$ similarity metric, other instances of $\textbf{S}$ and $M$ can be used depending on the structure under investigation.  For example, in references~\cite{zhenlidiamond} and~\cite{keys07}, we used different bond order parameters as shape descriptors to identify particles in the diamond lattice and in dodecagonal quasicrystals, respectively.  Arbitrarily complex crystal structures can be treated using this method by harmonic descriptors with a full spectrum of Fourier coefficients or Zernike moments.  We apply this scheme in the context of two examples highlighting the special symmetry properties of harmonic descriptors in section~\ref{sec:hoptricks} below.

\subsection{Database Search and Structure Identification}

In addition to computing order parameters and correlation functions, another common application of bond order parameters is to identify local structures such as icosahedral clusters\cite{gasser, keys07,iac07,gyroid} within a bulk system.  This typically involves choosing a cutoff value for a particular Fourier coefficient (for example, $|q_6|$) above which a cluster is identified as the structure of interest.  This is, in its essence, a rudimentary shape matching scheme, where the Fourier coefficient provides a coarse description of the cluster shape, and the cutoff acts as a similarity metric.  This type of structure identification scheme can be applied within a much broader context by using harmonic descriptors to perform a database search for an unknown structure.  The unknown structure is identified as the structure from the database of known structures that gives the best match.  Database searches based on harmonic descriptors have already been applied to proteins and macromolecules\cite{yeh, mak}.  In an earlier publication, we applied a database search to identify local structures within a phase separated system of tethered nanoparticles\cite{iac07}.  In the future, they may be applied to more abstract problems, such as data mining for web-accessible particle structures.

\begin{figure}
\begin{center}
\includegraphics[width=1.0\columnwidth]{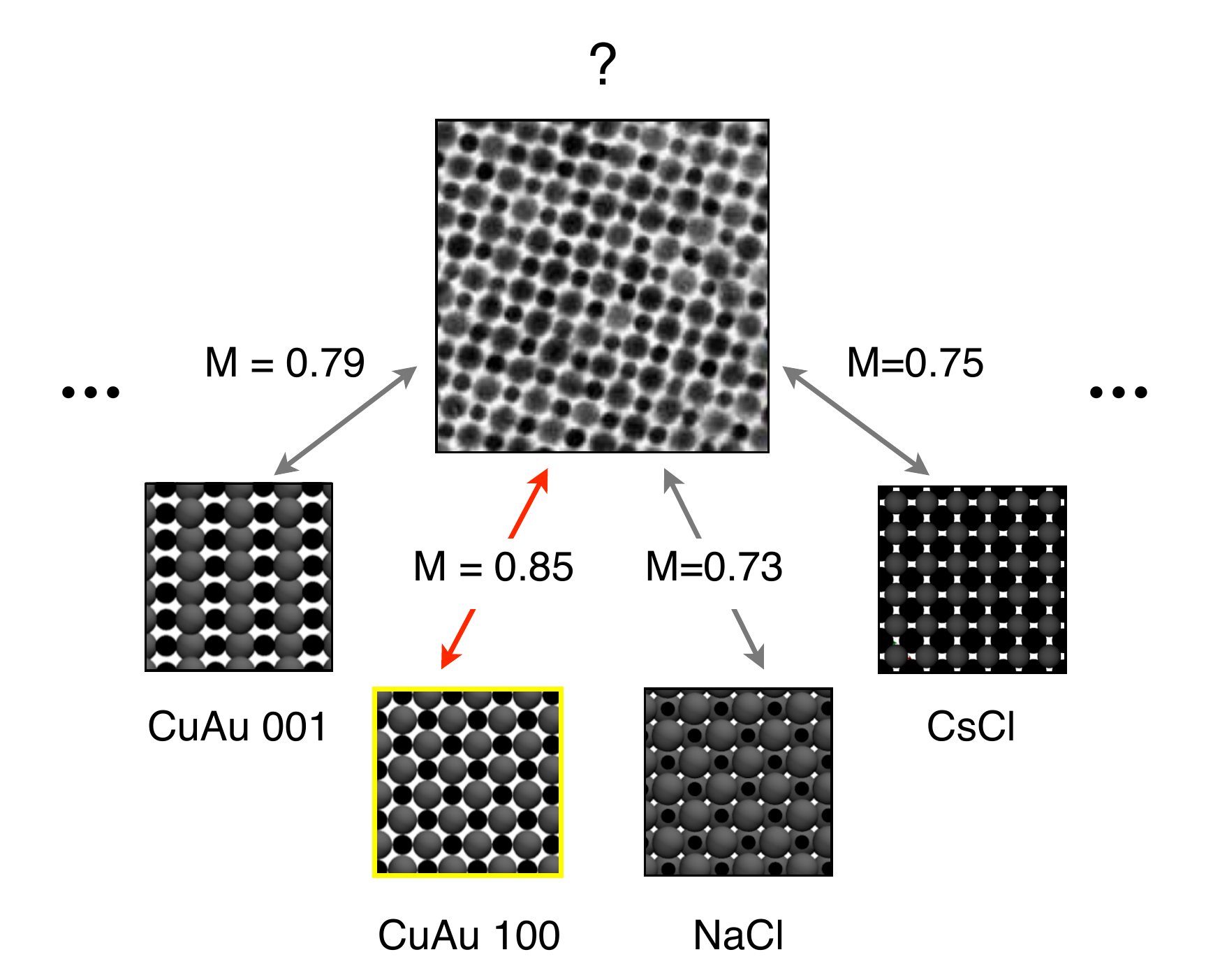}
\caption[Structure Identification Using a TEM Image] {Structure identification using a tunneling electron microscopy (TEM) image from reference~\cite{murray}.  The structure depicted in the image is identified by finding the best match from a database of candidate structures.  For this particular example, the reference structures are created by mathematical construction.  The image is correctly identified as depicting the CuAu structure viewed along the (100) direction.}
\label{fig:hop10}
\end{center}
\end{figure}

Database searches based on harmonic descriptors can be applied to a wide range of complex local and global structures.  As an example, consider the tunneling electron microscopy (TEM) image depicted in Fig.~\ref{fig:hop10}, which shows nanoparticles arranged in a binary crystalline superlattice from reference~\cite{murray}.  Although this structure was identified as the AuCu crystal viewed along the (100) direction by visual inspection\cite{murray}, assume for the purpose of this example that the structure is unknown.  The structure of the lattice can be identified by finding a best match from a database containing the images of known reference structures.  For our proof-of-concept example, we use a minimal reference database consisting of four different ideal binary crystal structures; however for more realistic problems, the database may be much more expansive.  The reference structures are created by mathematical construction and rendered by placing spheres at the lattice positions.  In practice, matching can be performed using other non-ideal images or other experimental images.  The images are indexed for comparison using the 2d Zernike descriptor $\textbf{S}^\textrm{Z2}$.  As mentioned previously, harmonic descriptors can be used to describe images\cite{zernike2d}, where $\{\textbf{x}\}$ and $\{f\}$ represent the positions and intensities of individual pixels. In practice, the pixel intensities are inverted ($I_i = -I_{i,0}$), since, for the current set of images, the particles are darker than the background.  To ensure that the matching algorithm is not affected by the different particle shades, we apply a binary thresholding criterion $I_i = \Theta(I_{i} - I_{cut})$.  To extract a global pattern from the image, we use the probability distributions method depicted in Fig.~\ref{fig:hop2}.  (Notice that although the structures are crystalline, the superposition method is not applicable, since the particle centers are not known).  For each local structure, we compute $\textbf{S}^\textrm{Z2}$ descriptors with rotation-invariant moments and frequencies in the range $0 \leq \ell \leq 8$.  Our overall results are not impacted by the inclusion of higher frequencies.  For each image, local descriptors are computed for 100 different randomly chosen pixels.  For each pixel, the range of neighboring pixels used to construct the local descriptor corresponds to roughly three particle diameters.  The local descriptors are then combined into an overall probability histogram descriptor for matching.  As shown in Fig.~\ref{fig:hop10}, the unknown structure most closely resembles the CuAu lattice along the (100) direction, in agreement with visual inspection\cite{murray}.   Additional orientations of the crystalline lattices could also be considered to better identify structures with complex ordering.

This same basic identification scheme  can be used for all types of structures, either simulated or experimental, local or global.  A subtlety arises when disordered local structures are possible, since it is typically infeasible to construct a reference database for the vast space of ``disordered'' structures.  As outlined in references~\cite{iac07, keys10-LONG} this problem can be avoided by setting a minimum value for the best match, below which structures are considered disordered.

In addition to database searches, experimental images can be used for all of the other applications described here or in references~\cite{keys10-ARCMP} and~\cite{keys10-LONG}, including quantifying structural perfection, computing order parameters and correlation functions, grouping and classifying structures, etc.  The only caveat is that, like simulation data, the images must be properly normalized such that the particle sizes and lattice spacings (if applicable) are the same for all compared structures.  In addition to images of particles, information can be extracted from other types of images, such as diffraction patterns.  For many applications, image processing can be applied to experimental images to simplify the data\cite{crocker1996, varadan2003, mohraz}, for example, by identifying particle centroids.  We explore the combination of image processing techniques and shape matching algorithms in a separate publication\cite{iac11preprint}.

\section{Special Properties of Harmonic Descriptors}
\label{sec:hoptricks}

In the previous section, we applied harmonic descriptors within the context of general particle shape matching applications, similar to those outlined in reference~\cite{keys10-LONG}.  Thus, although the harmonic descriptors have useful properties, such as rotation invariance, we could just as easily base our examples on other shape descriptors with similar properties.  In contrast, in this section we explore applications for which harmonic descriptors and Fourier coefficients are specifically well-suited, due to their unique symmetry properties.

\subsection{Matching to Within an n-Fold Rotation}

\begin{figure}
\begin{center}
\includegraphics[width=1.0\columnwidth]{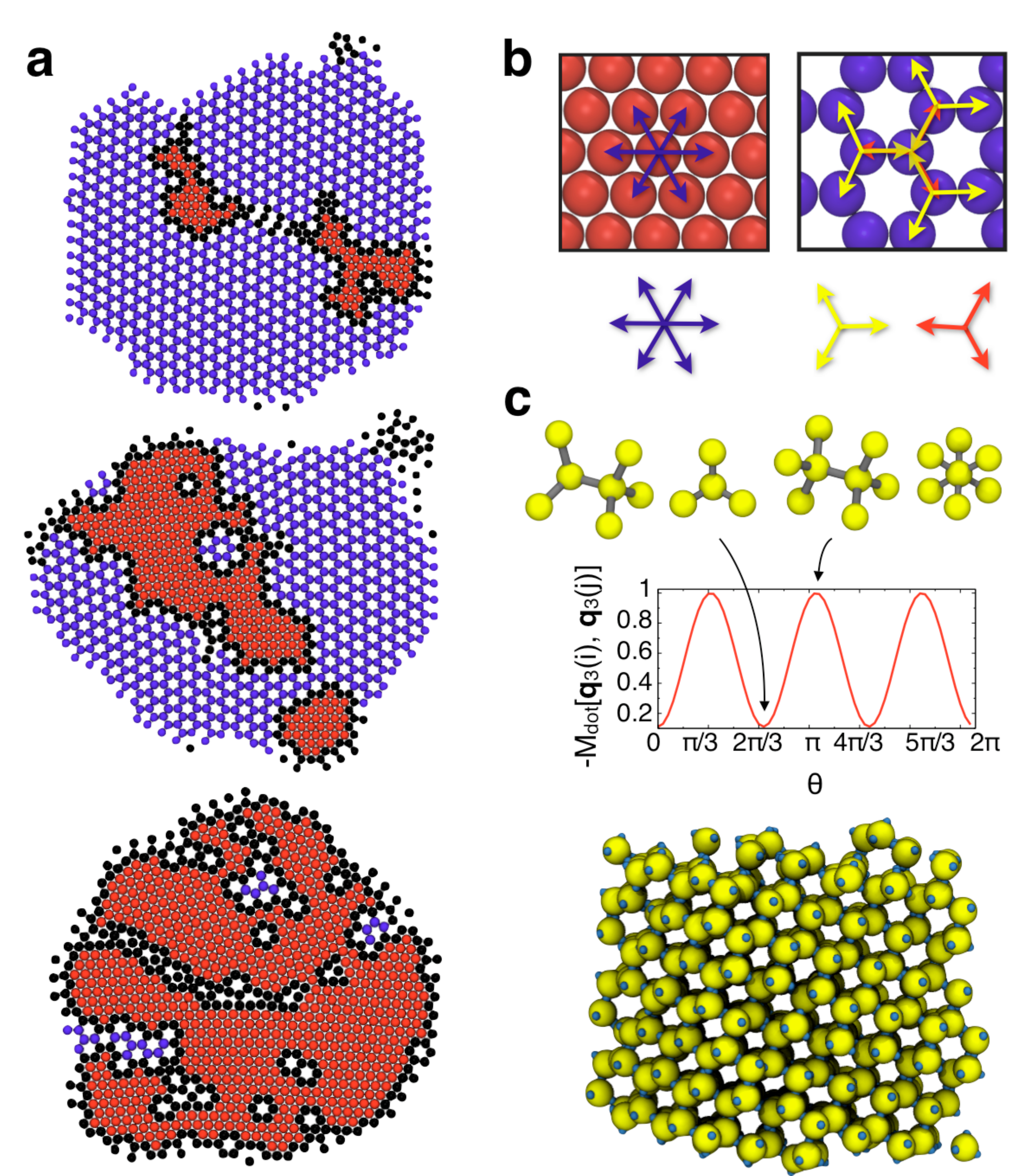}
\caption[Honeycomb to Hexagonal Transition in the 2d LJG system]{Honeycomb (hc) to hexagonal (hex) transition in the 2d Lennard-Jones Gauss (LJG) system\cite{engel07}.  (\textit{a}) Annealing results obtained for a LJG well-minimum position $r_0=1.57$ for three different values of the well depth parameter $\varepsilon$.  From top to bottom: $\varepsilon=2.7, 2.2, 1.4$.  Particles in the hc structure are colored blue, while particles in the hex structure are colored red.  All other particles are colored black.  (\textit{b}) Depiction of the local neighbor configurations in the hex and hc structures.  (\textit{c}) Diamond lattice formed by patchy particles\cite{zhenlidiamond, zhang2004}.  Analogously to the hc lattice, the diamond lattice can be characterized by the negated match of the third-order harmonic\cite{zhenlidiamond}.}
\label{fig:hoprot}
\end{center}
\end{figure}

One such application is the problem of matching structures that are a unique rotation of one another.  As an example, consider the problem of detecting the local crystal grains of different structures in the 2d system depicted in Fig.~\ref{fig:hoprot}a, where particles interact via the Lennard-Jones Gauss (LJG) potential\cite{engel07}.  For a potential-minimum-position parameter $r_0 = 1.57$, the system forms two crystal structures: the honeycomb (hc) structure at high values of the well-depth parameter $\varepsilon$ and the hexagonal (hex) structure at low $\varepsilon$\cite{engel07}.   As outlined in section~\ref{ssec:hopopcf}, local crystal-like pairs are typically identified by local clusters that match in terms of both shape and orientation.  This method is easily applicable to the hex lattice; however, in the case of the hc lattice, the triangular first neighbor shells of neighboring particles are mirrored and rotated by $60^\circ$ in the plane (see Fig.~\ref{fig:hoprot}b).  Thus, they match in terms of shape, but not in terms of orientation.  The symmetry properties of the Fourier coefficients pose a unique solution to this problem.  In the space of the $\ell=3$ Fourier coefficient $\psi_3$, the triangular neigbor shells in the hc lattice are precisely antiparallel (i.e., $M_{dot}[\psi_3(i),\ \psi(j)_3]= -1$, where $i$ and $j$ are neighbors).  Thus, a matching criterion can be constructed based on $-M_{dot}(\psi_{3},\ \psi_{3})$ to determine whether two neighbor shells are in the ideal hc configuration.  In Fig.~\ref{fig:hoprot}a, particles in the hc structure are colored blue, particles in the hex structure are colored red, and other particles, that do not belong to a particular crystal grain, are colored black.  The 3d analogy of this method, using $\textbf{q}_3$ in the place of $\psi_3$, was used to measure the number of particles in local diamond lattice grains in a system of patchy particles in reference~\cite{zhenlidiamond} (see Fig.~\ref{fig:hoprot}c).

\subsection{Matching Based on Rotational Symmetries}

\begin{figure}
\begin{center}
\includegraphics[width=1.0\columnwidth]{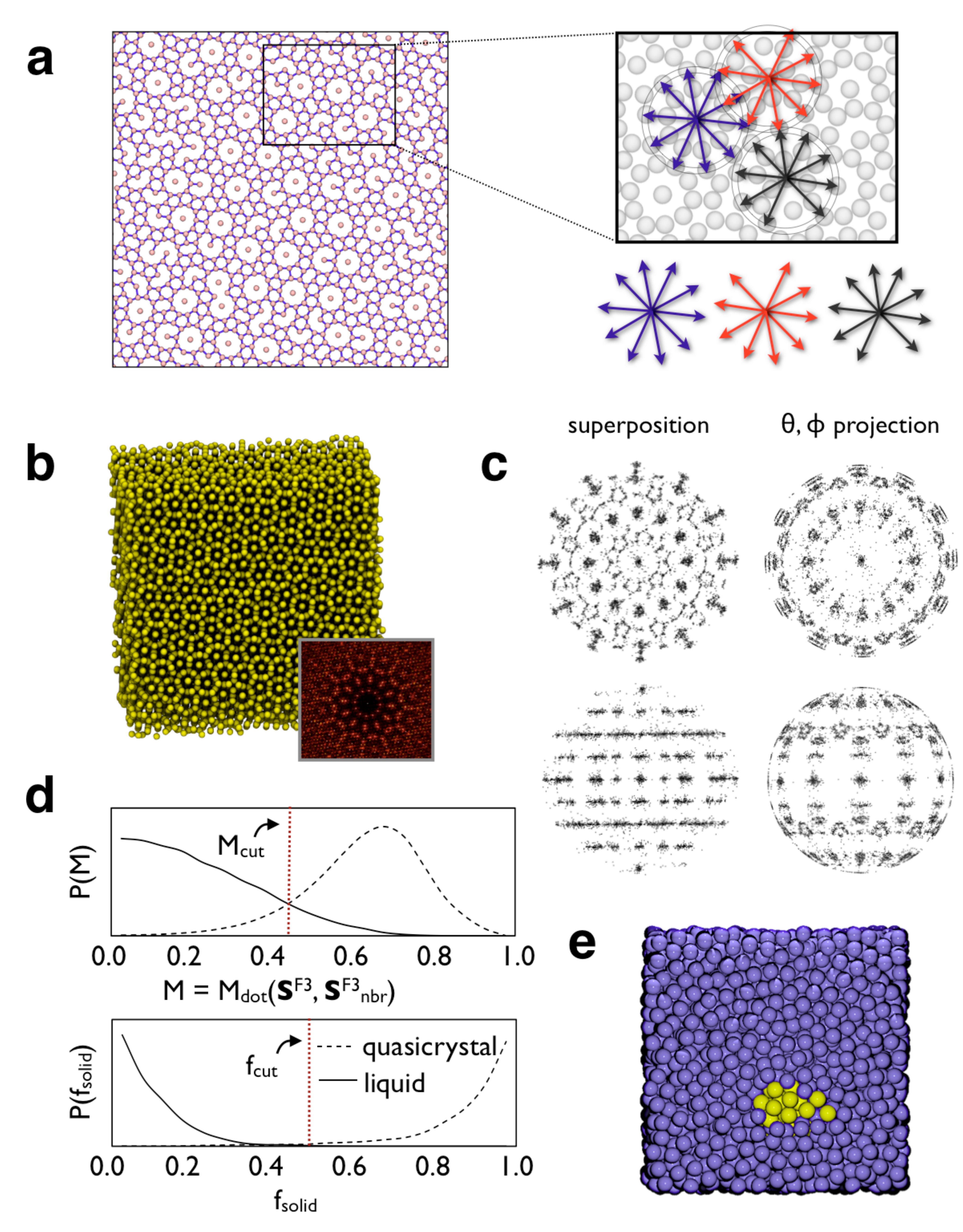}
\caption[Matching Based on Rotational Symmetries]{Matching based on rotational symmetries. (\textit{a}) Decagonal (10-fold) quasicrystal formed in the 2d LJG system\cite{engel07}.  The call-out depicts local structures within the system, which are 10-fold symmetric but dissimilar in shape.  (\textit{a}) Dodecagonal (12-fold) quasicrystal formed in the 3d Dzugutov system by MD simulation\cite{dzugutovdqc}, with a simulated diffraction image (lower right).  (\textit{b}) superposition of all neighbor clusters with a cutoff range 2.31$\sigma$.  The images on the right disregard $r$-dependence.  (\textit{c}) Probability distributions of bond correlations and solid-like neighbors used to determine the criterion for local quasicrystal grains.  (\textit{d})  Dzugutov liquid with a single small quasicrystal nucleus of about 50 atoms.}
\label{fig:hopsym}
\end{center}
\end{figure}

A similar application for which Fourier coeffcients are uniquely suited is the problem of matching structures based on their rotational symmetries rather than their shapes. As an example, consider the problem of matching local neighbor shells in the decagonal (i.e., 10-fold symmetric) quasicrystal formed in the 2d LJG system\cite{engel07} (Fig.~\ref{fig:hopsym}a).  Over the range indicated, the neighbor shells exhibit strong 10-fold rotational symmetry with a common direction, but the neighbor shells have different shapes.  Thus, our criterion for detecting local crystal grains outlined in section~\ref{ssec:hopopcf} fails for most shape descriptors, since the underlying shapes do not match.  As a solution, we can describe each local cluster with the $\ell=10$ Fourier coefficient $\psi_{10}$.  Since the clusters are 10-fold symmetric, and oriented in the same direction, the complex number $\psi_{10}$ is identical regardless of whether the clusters are missing particles.  Local quasicrystalline grains can then be detected as outlined in section~\ref{ssec:hopopcf}, using $M_{dot}(\psi_{10},\ \psi_{10})$ to identify local crystal-like pairs.  

In reference~\cite{keys07} we use an analogous scheme to detect ordered grains in a 3d dodecagonal (12-fold symmetric) quasicrystal (Fig.~\ref{fig:hopsym}b).  In this case, the structure has hundreds of different neighbor shell directions\cite{roth2000}, which exhibit strong 12-fold symmetry.  This is depicted by the superposition of local patterns over the range $r < 2.31\sigma$ (i.e., $\sim 2$ neighbor shells) in Fig.~\ref{fig:hopsym}c.  Considering this longer range ensures that each local region contains a sizable fraction of 12-fold directions.  Since we are only interested in the rotational symmetry and directionality of these local patterns, we remove the $r$-dependence from the patterns prior to matching (Fig.~\ref{fig:hopsym}c).  To capture the 12-fold symmetry, we use a matching criterion based on $M_{dot}(\textbf{q}_{12},\ \textbf{q}_{12})$.   Particles with a minimal fraction of solid-like matches $f_{cut}$ are considered to be locally quasicrystalline.  The cutoffs are determined by taking the crossover points in the probability distributions $P(M_{dot}(\textbf{q}_{12},\textbf{q}_{12}))$ and $P(f_{solid})$ (Fig.~\ref{fig:hopsym}d).  Following reference~\cite{keys07}, we take $M_{cut} = 0.45$, $f_{cut} = 0.5$.  As depicted in Fig.~\ref{fig:hopsym}e, this criterion is sufficient to determine a small quasicrystal nucleus in the bulk liquid. 

Notice that although both of our examples here are based on quasicrystalline structures, the method of matching dissimilar structures based on their rotational symmetries is applicable to a wide range of structures.  As a trivial example, we return to our previous problem of detecting crystal structures in the 2d LJG system.  Suppose now that we require an order parameter that detects crystalline grains in general, either hex or hc.  In this case, we can use a scheme based on the $\ell=6$ Fourier descriptor $\psi_6$, since in the space of $\ell=6$ harmonics, the triangle neighbor shells of the hc structure and the hexagon neighbor shells of the hex structure are equivalent.  

\subsection{Orientation About a Symmetry Axis}

\begin{figure*}
\begin{center}
\includegraphics[width=0.7\textwidth]{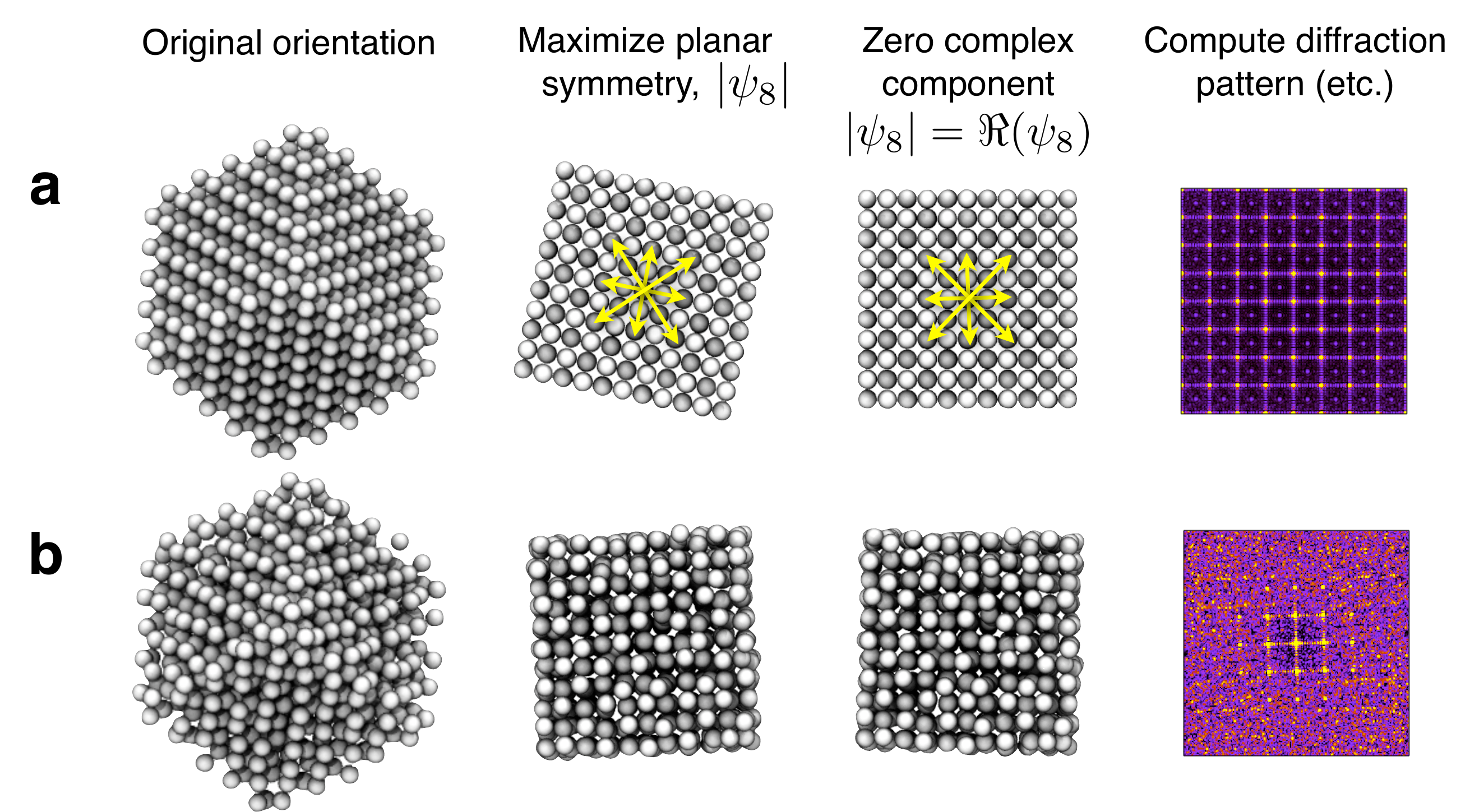}
\caption[Orientation About a Symmetry Axis] {Orientation about a symmetry axis.  (\textit{a}) Depiction of an algorithm for orienting a face-centered cubic (fcc) crystal about an 8-fold planar symmetry axis, in this case, for the purpose of automatically computing the diffraction image.  First, the planar symmetry, measured by $|\psi_8|$, is maximized such that the 8-fold symmetry axis is oriented in the z-direciton (out of the plane).  Then the structure is rotated in the xy plane to zero the complex component $\Im(\psi_8)$, such that the symmetry axis is oriented with the xy axis.  Finally, the diffraction image is computed.  (\textit{b}) The same process, repeated for a structure with thermal disorder.}
\label{fig:hop11}
\end{center}
\end{figure*}

As a final example of an application that exploits the symmetry properties of Fourier coefficients, consider the problem of aligning a crystal structure about a particular symmetry axis.  This is useful, for example, for computer algorithms that depend on the orientational direction of a structure, such as automatically computing the diffraction pattern or rendering images of particle data.  Suppose, for example that the desired symmetry axis is the $\ell = \ell_{0}$ plane, and the desired alignment direction is the z-axis.  The crystal can be iteratively rotated to maximize the Fourier coefficient $| \psi_{\ell_0} |$ where all particles are projected onto the xy plane.  If the crystal consists of a single grain, a relatively small test cluster can be used to perform the optimization, greatly enhancing computational efficiency.
 
As an example, consider the problem of aligning the face-centered cubic (fcc) crystal shown in Fig.~\ref{fig:hop11}a along an 8-fold planar symmetry axis.  Finding the optimal rotation that maximizes $| \psi_{8} |$ in the plane is solved by applying a simple simulated annealing Metropolis Monte Carlo (MC) scheme.  Our MC scheme involves attempting trial rotations of a small test cluster from the center of the box, which are accepted according to a Boltzmann probability distribution $\exp [ -\beta | \psi_{8} | ]$, with the fictitious energy function $-| \psi_{8} |$.  The inverse temperature $\beta$ is increased from 0.5 to 1000 over 100 steps.  In practice, the scheme only converges to a local minimum in the energy $-|\psi_{8} |$, so many different initial orientations are attempted to find a global minimum.  As depicted in Fig.~\ref{fig:hop11}, optimizing $| \psi_8 |$ aligns the 8-fold symmetry axis of the structure along the z-axis, but the structure is misaligned with the x and y-axes.  To perform this alignment, we zero the complex component $\Im(\psi_8)$, such that  $\Re(\psi_8) = | \psi_8 |$.  As depicted in Fig.~\ref{fig:hop11}b, this scheme is robust, even under a large amount of thermal noise.  More complex optimization algorithms may be applied to improve computational efficiency and accuracy over our simple MC scheme.  This type of orientation algorithm is potentially useful for matching diffraction data, since a large number of simulated diffraction images for 3d structures can easily be computed automatically about various symmetry axes.

\section{Summary and Future Outlook}

In summary, we have demonstrated how bond order parameters, already defined for particle structures on the unit circle and sphere, can be extended to index structures on the unit disk or ball.  We have demonstrated how these bond order parameters can be used to create harmonic shape descriptors, which can in turn be applied to create unique, highly specialized order parameters and automatically identify unknown particle structures.  In addition to the minimal proofs of concept reviewed here, more complex matching applications are explored in reference~\cite{keys10-LONG}.

In the future, the ability to describe structures numerically lends itself to many novel applications.  In the short term, matching applications can be used to automate structural analysis for large datasets.  Shape descriptors can be used within the context of many of the enhanced computational algorithms used in self-assembly and computational biology, such as path sampling\cite{tps, ffs} or metadynamics\cite{metadynamics} in the context of pseudo order parameters or collective variables to guide the sampling.  Shape descriptors can also serve as the basis for new optimization algorithms such as genetic algorithms\cite{genetic}, which often rely on energy rather than structure as a numerical measure of fitness.  In addition to abstract computational applications, shape matching algorithms can be applied to experimental images to obtain quantitative insight into experimental data.  By combining shape matching algorithms with new image processing schemes, much of the same information that we have obtained for simulation data can be obtained for experimental systems as well.


\textbf{Acknowledgements:} ASK was partially supported by a grant from the U.S. Department of Education (GAANN Grant No. P200A070538). SCG and ASK received partial support from the U.S. National Science Foundation (Grant Nos. DUE-0532831 and CHE-0626305). SCG and CRI received support from the U.S. Department of Energy, Office of Basic Energy Sciences, Division of Materials Sciences and Engineering under Award \#  DE-FG02-02ER46000.  CRI also acknowledges the University of Michigan Rackham Predoctoral Fellowship program. We thank T.D. Nguyen for helpful comments on the manuscript. Thanks also to T.D. Nguyen, M. Engel, A. Haji-Akbari, E.P. Jankowski, C. Phillips, D. Ortiz, A. Santos, I. Pons, and C. Singh for providing example data, not all of which could be used here.

\bibliography{References}

\begin{thebibliography}{10}

\bibitem{meyers2008}
M.~Meyers, P.~Chen, A.~Lin, and Y.~Seki, ``Biological materials: {S}tructure
  and mechanical properties,'' {\em Progress in Materials Science}, vol.~53,
  no.~1, pp.~1--206, 2008.

\bibitem{wang2005}
Y.~Wang, S.~Jian, S.~Han, S.~Feng, Z.~Feng, B.~Cheng, and D.~Zhang, ``Photonic
  band-gap engineering of quasiperiodic photonic crystals,'' {\em Journal of
  Applied Physics}, vol.~97, no.~10, p.~106112, 2005.

\bibitem{yoon2005}
J.~Yoon, W.~Lee, and E.~L. Thomas, ``Self-assembly of block copolymers for
  photonic-bandgap materials,'' {\em MRS Bulletin}, vol.~30, no.~10,
  pp.~721--726, 2005.

\bibitem{confinedfluids}
P.~T. Cummings, H.~Docherty, C.~R. Iacovella, and J.~K. Singh, ``Phase
  transitions in nanoconfined fluids: {T}he evidence from simulation and
  theory,'' {\em AIChE Journal}, vol.~56, no.~4, pp.~842--848, 2010.

\bibitem{murray}
E.~V. Shevchenko, D.~V. Talapin, N.~A. Kotov, S.~O'Brien, and C.~B. Murray,
  ``Structural diversity in binary nanoparticle superlattices,'' {\em Nature},
  vol.~439, pp.~55--59, Jan 2006.

\bibitem{kumacheva}
Z.~Nie, A.~Petukhova, and E.~Kumacheva, ``Properties and emerging applications
  of self-assembled structures made from inorganic nanoparticles,'' {\em Nature
  Nanotechnology}, vol.~5, no.~1, pp.~15--25, 2010.

\bibitem{glotzer07}
S.~C. Glotzer and M.~J. Solomon, ``Anisotropy of building blocks and their
  assembly into complex structures,'' {\em Nature Materials}, vol.~6,
  pp.~557--562, 2007.

\bibitem{halperin78}
B.~I. Halperin and D.~R. Nelson, ``Theory of two-dimensional melting,'' {\em
  Physical Review Letters}, vol.~41, no.~2, pp.~121--124, 1978.

\bibitem{snr83}
P.~J. Steinhardt, D.~R. Nelson, and M.~Ronchetti, ``Bond-orientational order in
  liquids and glasses,'' {\em Physical Review B}, vol.~28, no.~2, pp.~784--805,
  1983.

\bibitem{gasser}
U.~Gasser, A.~Schofield, and D.~A. Weitz, ``Local order in a supercooled
  colloidal fluid observed by confocal microscopy,'' {\em Journal of Physics:
  Condensed Matter}, vol.~15, p.~S375, 2003.

\bibitem{iac07}
C.~R. Iacovella, A.~S. Keys, M.~A. Horsch, and S.~C. Glotzer, ``Icosahedral
  packing of polymer-tethered nanospheres and stabilization of the gyroid
  phase,'' {\em Physical Review E}, vol.~75, p.~040801(R), 2007.

\bibitem{marcus96}
A.~H. Marcus and S.~A. Rice, ``{Observations of first-order liquid-to-hexatic
  and hexatic-to-solid phase transitions in a confined colloid suspension},''
  {\em Physical Review Letters}, vol.~77, no.~12, pp.~2577--2580, 1996.

\bibitem{kawasaki}
T.~Kawasaki, T.~Araki, and H.~Tanaka, ``Correlation between dynamic
  heterogeneity and medium-range order in two-dimensional glass-forming
  liquids,'' {\em Physical Review Letters}, vol.~99, no.~21, p.~215701, 2007.

\bibitem{ernstnagelgrest}
R.~M. Ernst, S.~R. Nagel, and G.~S. Grest, ``Search for a correlation length in
  a simulation of the glass transition,'' {\em Physical Review B}, vol.~43,
  no.~10, pp.~8070--8080, 1991.

\bibitem{tenwolde96}
P.~R. ten Wolde, M.~J. Ruiz-Montero, and D.~Frenkel, ``{Numerical calculation
  of the rate of crystal nucleation in a Lennard-Jones system at moderate
  undercooling},'' {\em The Journal of Chemical Physics}, vol.~104,
  pp.~9932--9947, Jun 1996.

\bibitem{auer04}
S.~Auer and D.~Frenkel, ``Numerical prediction of absolute crystallization
  rates in hard-sphere colloids,'' {\em The Journal of Chemical Physics},
  vol.~120, pp.~3015--3029, Feb 2004.

\bibitem{gasser01}
U.~Gasser, E.~R. Weeks, A.~Schofield, P.~N. Pusey, and D.~A. Weitz,
  ``Real-space imaging of nucleation and growth in colloidal crystallization,''
  {\em Science}, vol.~292, no.~5515, p.~258, 2001.

\bibitem{tanaka}
H.~Shintani and H.~Tanaka, ``Frustration on the way to crystallization in
  glass,'' {\em Nature Physics}, vol.~2, no.~3, pp.~200--206, 2006.

\bibitem{keys07}
A.~S. Keys and S.~C. Glotzer, ``How do quasicrystals grow?,'' {\em Physical
  Review Letters}, vol.~99, no.~23, p.~235503, 2007.

\bibitem{tesfuv2}
T.~Solomon and M.~J. Solomon, ``Stacking fault structure in shear-induced
  colloidal crystallization,'' {\em The Journal of Chemical Physics}, vol.~124,
  no.~13, p.~134905, 2006.

\bibitem{glotzer2005}
S.~C. Glotzer, M.~A. Horsch, C.~R. Iacovella, Z.~Zhang, E.~R. Chan, and
  X.~Zhang, ``Self-assembly of anisotropic tethered nanoparticle shape
  amphiphiles,'' {\em Current Opinion in Colloid and Interface Science},
  vol.~10, pp.~287--295, Dec 2005.

\bibitem{desimone2006}
L.~E. Euliss, J.~A. DuPont, S.~Gratton, and J.~DeSimone, ``Imparting size,
  shape, and composition control of materials for nanomedicine,'' {\em Chemical
  Society Reviews}, vol.~35, no.~11, pp.~1095--1104, 2006.

\bibitem{amir09}
A.~Haji-Akbari, M.~Engel, A.~S. Keys, X.~Zheng, R.~G. Petschek,
  P.~Palffy-Muhoray, and S.~C. Glotzer, ``Disordered, quasicrystalline and
  crystalline phases of densely packed tetrahedra,'' {\em Nature}, vol.~462,
  no.~7274, pp.~773--777, 2009.

\bibitem{tang2006}
Z.~Y. Tang, Z.~Zhang, Y.~Wang, S.~C. Glotzer, and N.~A. Kotov, ``{Self-assembly
  of CdTe nanocrystals into free-floating sheets},'' {\em Science}, vol.~314,
  pp.~274--278, Oct 2006.

\bibitem{zhang2007}
Z.~Zhang, Z.~Y. Tang, N.~A. Kotov, and S.~C. Glotzer, ``{Simulations and
  analysis of self-assembly of CdTe nanoparticles into wires and sheets},''
  {\em Nano Letters}, vol.~7, pp.~1670--1675, Jun 2007.

\bibitem{zhang2003}
Z.~Zhang, M.~A. Horsch, M.~H. Lamm, and S.~C. Glotzer, ``Tethered nano building
  blocks: {T}oward a conceptual framework for nanoparticle self-assembly,''
  {\em Nano Letters}, vol.~3, no.~10, pp.~1341--1346, 2003.

\bibitem{park}
J.-W. Park and E.~L. Thomas, ``Anisotropic micellar nanoobjects from reactive
  liquid crystalline rod−coil diblock copolymers,'' {\em Macromolecules},
  vol.~37, no.~10, pp.~3532--3535, 2004.

\bibitem{reister}
E.~Reister and G.~H. Fredrickson, ``Phase behavior of a blend of
  polymer-tethered nanoparticles with diblock copolymers,'' {\em The Journal of
  Chemical Physics}, vol.~123, no.~21, p.~214903, 2005.

\bibitem{arthi2008}
A.~Jayaraman and K.~S. Schweizer, ``Structure and assembly of dense solutions
  and melts of single tethered nanoparticles,'' {\em The Journal of Chemical
  Physics}, vol.~128, no.~16, 2008.

\bibitem{waddon2002}
A.~J. Waddon, L.~Zheng, R.~J. Farris, and E.~B. Coughlin, ``{Nanostructured
  polyethylene-POSS copolymers: Control of crystallization and aggregation},''
  {\em Nano Letters}, vol.~2, pp.~1149--1155, Oct 2002.

\bibitem{rotello}
A.~K. Boal, F.~Ilhan, J.~E. DeRouchey, T.~Thurn-Albrecht, T.~P. Russell, and
  V.~M. Rotello, ``Self-assembly of nanoparticles into structured spherical and
  network aggregates,'' {\em Nature}, vol.~404, no.~6779, pp.~746--748, 2000.

\bibitem{virus}
T.~Chen, Z.~Zhang, and S.~C. Glotzer, ``A precise packing sequence for
  self-assembled convex structures,'' {\em Proceedings of the National Academy
  of Sciences}, vol.~104, no.~3, pp.~717--722, 2007.

\bibitem{hagan2006}
M.~F. Hagan and D.~Chandler, ``{Dynamic pathways for viral capsid assembly},''
  {\em Biophysical Journal}, vol.~91, no.~1, pp.~42--54, 2006.

\bibitem{keys10-ARCMP}
A.~S. Keys, C.~R. Iacovella, and S.~C. Glotzer, ``Characterizing structure
  through shape matching and applications to self-assembly,'' {\em Annual
  Reviews of Condensed Matter Physics}, vol.~2, 2011.

\bibitem{keys10-LONG}
A.~S. Keys, C.~R. Iacovella, and S.~C. Glotzer, ``Characterizing structure in
  assembled systems using shape matching : {A}lgorithms and applications.''
  preprint, 2010.

\bibitem{yeh}
J.~S. Yeh, D.~Y. Chen, B.~Y. Chen, and M.~Ouhyoung, ``A web-based
  three-dimensional protein retrieval system by matching visual similarity,''
  {\em Bioinformatics}, vol.~21, no.~13, p.~3056, 2005.

\bibitem{mak}
L.~Mak, S.~Grandison, and R.~J. Morris, ``An extension of spherical harmonics
  to region-based rotationally invariant descriptors for molecular shape
  description and comparison,'' {\em Journal of Molecular Graphics and
  Modelling}, vol.~26, no.~7, pp.~1035--1045, 2008.

\bibitem{venkatraman}
V.~Venkatraman, L.~Sael, and D.~Kihara, ``{Potential for protein surface shape
  analysis using spherical harmonics and 3D Zernike descriptors},'' {\em Cell
  Biochemistry and Biophysics}, vol.~54, no.~1, pp.~23--32, 2009.

\bibitem{smwebsite}
A.~S. Keys and C.~R. Iacovella, ``Particle shape matching library and
  examples,'' {\em URL http://www. glotzerlab. engin. umich. edu/shapematching.
  html}, 2010.

\bibitem{horsch2005}
M.~A. Horsch, Z.~Zhang, and S.~C. Glotzer, ``Self-assembly of polymer-tethered
  nanorods,'' {\em Physical Review Letters}, vol.~95, no.~5, p.~056105, 2005.
\newblock 056105.

\bibitem{gyroid}
C.~R. Iacovella, M.~A. Horsch, and S.~C. Glotzer, ``{Local ordering of
  polymer-tethered nanospheres and nanorods and the stabilization of the double
  gyroid phase},'' {\em The Journal of Chemical Physics}, vol.~129, p.~044902,
  2008.

\bibitem{roth2000}
J.~Roth and A.~R. Denton, ``{Solid-phase structures of the Dzugutov pair
  potential},'' {\em Physical Review E}, vol.~61, no.~6, pp.~6845--6857, 2000.

\bibitem{zhenlidiamond}
Z.~Zhang, A.~S. Keys, T.~Chen, and S.~C. Glotzer, ``Self-assembly of patchy
  particles into diamond structures through molecular mimicry,'' {\em
  Langmuir}, vol.~21, no.~25, pp.~11547--11551, 2005.

\bibitem{ungar03}
G.~Ungar, Y.~Liu, X.~Zeng, V.~Percec, and W.~D. Cho, ``Giant supramolecular
  liquid crystal lattice,'' {\em Science}, vol.~299, no.~5610, pp.~1208--1211,
  2003.

\bibitem{tnv}
T.~D. Nguyen, Z.~Zhang, and S.~C. Glotzer, ``Molecular simulation study of
  self-assembly of tethered v-shaped nanoparticles,'' {\em The Journal of
  Chemical Physics}, vol.~129, p.~244903, 2008.

\bibitem{tns}
C.~R. Iacovella, M.~A. Horsch, Z.~Zhang, and S.~C. Glotzer, ``Phase diagrams of
  self-assembled mono-tethered nanospheres from molecular simulation and
  comparison to surfactants,'' {\em Langmuir}, vol.~21, no.~21, pp.~9488--9494,
  2005.

\bibitem{ditethered}
C.~R. Iacovella and S.~C. Glotzer, ``Complex crystal structures formed by the
  self-assembly of ditethered nanospheres,'' {\em Nano Letters}, vol.~9, no.~3,
  pp.~1206--1211, 2009.

\bibitem{hungarian}
H.~W. Kuhn, ``{The Hungarian method for the assignment problem},'' {\em Naval
  Research Logistics Quarterly}, vol.~2, no.~1-2, pp.~83--97, 1955.

\bibitem{ankerst}
M.~Ankerst, G.~Kastenmuller, H.~P. Kriegel, and T.~Seidl, ``3d shape histograms
  for similarity search and classification in spatial databases,'' {\em Lecture
  Notes in Computer Science}, pp.~207--228, 1999.

\bibitem{icp}
P.~J. Besl and H.~D. McKay, ``{A method for registration of 3-D shapes},'' {\em
  IEEE Transactions on Pattern Analysis and Machine Intelligence}, vol.~14,
  no.~2, pp.~239--256, 1992.

\bibitem{pca}
G.~H. Dunteman, {\em {Principal Components Analysis}}.
\newblock Sage Publications, Inc., Newbury Park, CA, U.S.A., 1989.

\bibitem{ylm}
M.~Kazhdan, T.~Funkhouser, and S.~Rusinkiewicz, ``Rotation invariant spherical
  harmonic representation of 3d shape descriptors,'' in {\em Proceedings of the
  2003 Eurographics/ACM SIGGRAPH Symposium on Geometry Processing}, p.~164,
  Eurographics Association, 2003.

\bibitem{fourier}
C.~T. Zahn and R.~Z. Roskies, ``Fourier descriptors for plane closed curves,''
  {\em IEEE Transactions on Computers}, vol.~21, no.~3, pp.~269--281, 1972.

\bibitem{ylm2}
D.~V. Vranic, D.~Saupe, and J.~Richter, ``{Tools for 3D-object retrieval:
  Karhunen-Loeve transform and spherical harmonics},'' in {\em IEEE 2001
  Workshop Multimedia Signal Processing}, pp.~293--298, Citeseer, 2001.

\bibitem{gubbins}
R.~Radhakrishnan, K.~E. Gubbins, and M.~Sliwinska-Bartkowiak, ``Effect of the
  fluid-wall interaction on freezing of confined fluids: Toward the development
  of a global phase diagram,'' {\em The Journal of Chemical Physics}, vol.~112,
  no.~24, pp.~11048--11057, 2000.

\bibitem{fsht}
M.~J. Mohlenkamp, ``A fast transform for spherical harmonics,'' {\em Journal of
  Fourier Analysis and Applications}, vol.~5, no.~2, pp.~159--184, 1999.

\bibitem{fft}
G.~D. Bergland, ``{A guided tour of the fast Fourier transform},'' {\em IEEE
  Spectrum}, vol.~6, no.~7, pp.~41--52, 2009.

\bibitem{iacovella2009b}
C.~R. Iacovella and S.~C. Glotzer, ``Phase behavior of ditethered
  nanospheres,'' {\em Soft Matter}, vol.~5, pp.~4492 -- 4498, 2009.

\bibitem{zernike2d}
A.~Khotanzad and Y.~H. Hong, ``{Invariant image recognition by Zernike
  moments},'' {\em IEEE Transactions on Pattern Analysis and Machine
  Intelligence}, vol.~12, no.~5, pp.~489--497, 1990.

\bibitem{zernike3d}
M.~Novotni and R.~Klein, ``{3D Zernike descriptors for content based shape
  retrieval},'' in {\em Proceedings of the Eighth ACM Symposium on Solid
  Modeling and Applications}, pp.~216--225, ACM New York, N. Y., U. S.A, 2003.

\bibitem{ubiquitin}
A.~Hershko and A.~Ciechanover, ``{The Ubiquitin system},'' {\em Annual Review
  of Biochemistry}, vol.~67, no.~1, pp.~425--479, 1998.

\bibitem{namd}
J.~C. Phillips, R.~Braun, W.~Wang, J.~Gumbart, E.~Tajkhorshid, E.~Villa,
  C.~Chipot, R.~D. Skeel, L.~Kale, and K.~Schulten, ``{Scalable molecular
  dynamics with NAMD},'' {\em Journal of Computational Chemistry}, vol.~26,
  no.~16, p.~1781, 2005.

\bibitem{vmd}
W.~Humphrey, A.~Dalke, and K.~Schulten, ``{VMD: Visual molecular dynamics},''
  {\em Journal of Molecular Graphics}, vol.~14, no.~1, pp.~33--38, 1996.

\bibitem{nelsonc6}
D.~R. Nelson and B.~I. Halperin, ``Dislocation-mediated melting in two
  dimensions,'' {\em Physical Review B}, vol.~19, no.~5, pp.~2457--2484, 1979.

\bibitem{laura}
L.~T. Shereda, R.~G. Larson, and M.~J. Solomon, ``Local stress control of
  spatiotemporal ordering of colloidal crystals in complex flows,'' {\em
  Physical Review Letters}, vol.~101, no.~3, p.~38301, 2008.

\bibitem{crocker1996}
J.~C. Crocker and D.~G. Grier, ``Methods of digital video microscopy for
  colloidal studies,'' {\em Journal of Colloid and Interface Science},
  vol.~179, pp.~298--310, Apr 1996.

\bibitem{varadan2003}
P.~Varadan and M.~J. Solomon, ``Direct visualization of long-range
  heterogeneous structure in dense colloidal gels,'' {\em Langmuir}, vol.~19,
  no.~3, pp.~509--512, 2003.

\bibitem{mohraz}
A.~Mohraz and M.~J. Solomon, ``{Direct visualization of colloidal rod assembly
  by confocal microscopy},'' {\em Langmuir}, vol.~21, no.~12, pp.~5298--5306,
  2005.

\bibitem{iac11preprint}
C.~R. Iacovella, A.~S. Keys, and S.~C. Glotzer, ``Image processing techniques
  for structural analysis of particle systems.'' preprint, 2011.

\bibitem{engel07}
M.~Engel and H.~R. Trebin, ``Self-assembly of monatomic complex crystals and
  quasicrystals with a double-well interaction potential,'' {\em Physical
  Review Letters}, vol.~98, no.~22, p.~225505, 2007.

\bibitem{zhang2004}
Z.~Zhang and S.~C. Glotzer, ``Self-assembly of patchy particles,'' {\em Nano
  Letters}, vol.~4, pp.~1407--1413, Aug 2004.

\bibitem{dzugutovdqc}
M.~Dzugutov, ``Formation of a dodecagonal quasicrystalline phase in a simple
  monatomic liquid,'' {\em Physical Review Letters}, vol.~70, no.~19,
  pp.~2924--2927, 1993.

\bibitem{tps}
C.~Dellago, P.~G. Bolhuis, F.~S. Csajka, and D.~Chandler, ``Transition path
  sampling and the calculation of rate constants,'' {\em The Journal of
  Chemical Physics}, vol.~108, p.~1964, 1998.

\bibitem{ffs}
R.~J. Allen, D.~Frenkel, and P.~R. ten Wolde, ``Forward flux sampling-type
  schemes for simulating rare events: {E}fficiency analysis,'' {\em The Journal
  of Chemical Physics}, vol.~124, p.~194111, 2006.

\bibitem{metadynamics}
A.~Laio and M.~Parrinello, ``Escaping free-energy minima,'' {\em Proceedings of
  the National Academy of Sciences}, vol.~99, no.~20, pp.~12562--12566, 2002.

\bibitem{genetic}
D.~E. Goldberg, {\em {Genetic Algorithms in Search, Optimization, and Machine
  Learning}}.
\newblock Addison-Wesley Reading Menlo Park, 1989.

\end{thebibliography}
\bibliographystyle{ieeetr}

\end{document}